\documentclass[]{aa}
\usepackage{graphicx}
\usepackage{txfonts}
\usepackage{lscape}
\usepackage{rotating}
\usepackage{natbib}
\usepackage{aalongtable}

\def\etal{et al. }

\def\Ia{SN~Ia~}

\begin{document}
\def\lesim{\stackrel{<}{{}_{\sim}}} \title{Spectroscopic observations of eight supernovae at intermediate redshift} \author{C.  Balland\inst{1,2,3}, M. Mouchet\inst{3,4}, 
R. Amanullah\inst{5}, 
P. Astier\inst{1}, S.  Fabbro\inst{6}, 
G. Folatelli\inst{5}, G. Garavini\inst{5},
A. Goobar\inst{5}, D. Hardin\inst{1}, M. J. Irwin\inst{7}, R. G. McMahon\inst{7}, A.-M., Mour\~ao\inst{6}, S. Nobili\inst{5}, R. Pain\inst{1}, R. Pascoal\inst{6}, J. Raux\inst{1}, G. Sainton\inst{1}, K. Schahmaneche\inst{1}, 
N. A. Walton\inst{7}}

\institute{LPNHE, CNRS-IN2P3 and Universities of Paris 6 \& 7,
F-75252 Paris Cedex 05, France 
\and Univ. Paris-Sud, F-91405 Orsay, France
\and APC, UMR 7164 CNRS, 11 Place Marcelin Berthelot, F-75231 Paris Cedex 05, France
\and LUTH, UMR 8102 CNRS, Observatoire de Paris, Section de Meudon, F-92195 Meudon Cedex, France
\and  Department of Physics, Stockholm University, SE-10691 Stockholm, Sweden
\and CENTRA-Centro M. de Astrofisica and Department of Physics, IST, Lisbon, Portugal
\and Institute of Astronomy, University of Cambridge, Madingley Road,
Cambridge, CB3 0HA, UK
}

\offprints{christophe.balland@ias.u-psud.fr}

\date{Received; accepted} \titlerunning{Spectroscopic observations of eight
  supernovae at intermediate redshifts}
\authorrunning{Balland \etal}

\abstract
  {}
  {We present spectra of six Type Ia and two Type II supernovae
  obtained in June 2002 at the William Herschel Telescope during a
  search for Type Ia supernovae (SN~Ia) at intermediate redshift. }
  {Supernova type identification and phase determination are performed using a fitting technique based on a $\chi^2$
  minimization against a series of model templates.}
  { The
  spectra range from $z=0.033$ to $z=0.328$, including one
  spectroscopically underluminous \Ia at $z=0.033$.  This set of
  spectra significantly increases the sample of well-observed type \Ia
  supernovae available in the range $0.15\lesim z\lesim 0.35$. 
  Together with the twelve supernovae observed by our team in 1999 in
  the same redshift range, they form an homogeneous sample of
  seventeen type Ia supernovae with comparable signal-to-noise ratio
  and regular phase sampling in a still largely unexplored region of
  the redshift space.}
  {}

\maketitle

\section{Introduction}

Spectroscopic observations of type Ia supernovae (SN~Ia) at various
redshifts are crucial when building the Hubble diagram from which
cosmological parameters are derived. They are necessary to confirm the
nature and type of the candidates, which are usually selected by
photometric means. They also provide the only way, so far known, to
measure redshifts with sufficient accuracy to be used in constructing
the Hubble diagram.  Moreover, spectra give clues to the diversity of
type Ia supernovae \citep{Benetti05,Hachinger06} and allow one to
test the fundamental hypothesis underlying these cosmological studies:
that type Ia supernovae are standardizable candles.

Much effort has thus been put in the last decade to detect and follow
up SN~Ia, both photometrically and spectroscopically, at high redshifts
where the cosmological test is most significant and at low redshifts
where observations are necessary to anchor the Hubble diagram.
Spectroscopic and photometric programs have been scheduled in parallel
as part of unified observational programs on various telescopes during
the pioneering SN searches in the late 1990s and early 2000s. More
recently, these programs are typically scheduled in service observing
mode at the 8-10m class telescopes in second generation, larger scale
searches such as the ones undertaken by the SNLS \citep{Astier06} or
the ESSENCE/High-Z teams \citep{Barris04,Matheson05}. Whereas the high
redshift supernova searches have gathered several hundreds spectra of
type Ia supernovae \citep{Basa06,Howell05,Hook05} and spectroscopic
programs are underway to study low redshift \Ia (LOSS;
  \citet{Filippenko01}, SNFactory; \citet{Aldering02}), the redshift
range $0.15\lesim z \lesim 0.3$ has been relatively unexplored so far.
This situation is about to change with the forthcoming publication of
the Sloan Digital Sky Survey (SDSS; \citet{Sako05b,Sako05a}) supernova
sample at intermediate redshift. This should provide an unprecedented
collection of supernova spectra at these redshifts. Meanwhile, smaller
scale searches of \Ia at intermediate redshifts have been undertaken
by various groups, e.g. \citet{Riello03}, \citet{Balland06}.  The
spectra described in this paper are part of such pioneering
observations.

Studying \Ia at intermediate redshifts is of crucial importance. These
supernovae not only allow for the filling of the gap observed in the
current Hubble diagram, but also their spectra allow for studying the
evolution of supernova properties with redshift as well as the
evolution of the rate of peculiar (under- or overluminous)
supernovae that can pollute supernova samples and thus potentially
alter the derived cosmological results. There is growing evidence that
diversity rather than uniformity is common among type Ia supernovae at low
redshifts and it is important to know whether this diversity is also
found at higher redshifts and how it may limit the precision on the
cosmological parameters one can expect from this cosmological test
based on the Hubble diagram.

For these reasons, a systematic search for \Ia at intermediate
redshift was carried out at the Isaac Newton Group of Telescopes on La
Palma Observatory in 2002 as part of the International Time Program
(ITP2002; Ruiz-Lapuente et al.)  on "{\em $\Omega$ and $\Lambda$ from
  Supernovae, and the Physics of Supernova Explosions}".  The spectra
presented in this paper were obtained on June 10th and 11th, 2002
at the William Herschel Telescope (WHT).  Photometric detection and
follow-up were obtained at the 2.5m Isaac Newton Telescope, the
1.0m Jakobus Kapteyn Telescope and the Nordic Optical telescope.
Preliminary results of this search have been reported by
\citet{Altavilla05}.  The present paper presents in detail the fully
reduced spectra and their analysis.  In Sect.~2, we briefly summarize
the detection strategy. More details can be found in \citet{Balland06}
as the strategy described in that paper is by many respects comparable
to the one adopted in this search.  In Sect.~3, we present the
spectroscopic observations and reduction techniques. We present the
spectroscopic analysis and results in Sects.~4 and 5. Finally, we
discuss our spectra and conclude in Sect.~6.

\section{Strategy for discovering  \Ia at intermediate redshift}
 
The search for supernovae used the INT Prime Focus Wide Field Camera
(WFC) with its array of four 4096$\times$2048 thinned EEV CCD's. This
camera has a $\sim 0.29$ square degree FOV. Repeated observations of
the same area of the sky were obtained for consecutive 4 nights (June
4th to June 7th 2002) in one photometric band (g') on 88 separate
fields. These were selected from the Wide Field Survey (WFS) fields,
including the 1610+5430 Elais N1, 1637+4117 Elais N2 and 2240+0000
regions. Due to overlap of fields, about 20 square degrees were
surveyed over the discovery period, for a total of 2 exposures of 480
seconds with an average seeing of $\approx 1$~arcsec. The typical
depth achieved was g'$\approx 25$ \citep{Walton99,McMahon01}.
Subtraction images were then obtained by subtracting a reference image
of the same field obtained two or three years before (these reference
images were g' band 600s exposures), from either the first, the second
or the sum of two new exposures.  During the first three nights, the
search for point-like sources was performed automatically on the three
subtracted images, by requiring a positive subtraction simultaneously
on the three images and at the same location within two pixels. Cosmic
rays hits and transient objects are rejected by this procedure. The
candidates were then checked visually by several (at least two)
independent scanners. The selected candidates were then reobserved the
following night and kept only if present on these new images with a
similar or higher flux. This aims to limit the number of
  declining supernovae selected by our procedure.  Finally, supernova
candidates were classified as very likely, probable or possible
supernovae according to their flux and shape properties, and
prioritized for spectroscopy. Applying this search procedure generated
25 supernova candidates over the 4 nights of the discovery period.
Seventeen candidates were sent to WHT for spectroscopy within a few
days after discovery, among which eight were subsequently confirmed as
supernovae \citep{Ruiz-Lapuente05}. The remaining candidates were
identified as various active galaxies ranging from $z=0.257$ to
$z=2.290$, and are not presented in this paper.

\section{Spectroscopic observations and reduction}

\subsection{Observations}

We used the two arms (red and blue) of the ISIS spectrometer to
obtain long-slit spectra of the supernova candidates previously
selected. Observations were carried out on June 10th and June 11th
2002. The weather was good during these two nights, with an average
seeing of $\approx 1$ arcsec.  For each candidate, except two (SN~2002lk,
SN~2002li), two exposures of 900s were obtained producing spectra over
the spectral range $3200-9000\AA$, using the R158B grating in the blue
arm and the R158R grating in the red arm, with the $6100\AA\ $
dichroic filter. The typical spectral resolution is $7-8.5\AA\ $ in
the blue and red arms, with fringing arising in the red spectra beyond
$\approx 7000\AA$.  A slit of $1.2$ arcsec width was used on June 10th
at the parallactic angle. On June 11th, the slit width was reduced to
1.0 arcsec, comparable to the seeing, except for one spectrum of
SN~2002lk and one of SN~2002lj.  Good quality spectra, with
signal-to-noise ratio greater than or of the order of 5 per pixel ($3
\AA\ $ bin), were obtained in this way. Spectra of two
spectrophotometric standard stars (BD+17 4708 and HZ 44) were acquired
at the beginning and at the end of each night and used to flux
calibrate the supernova spectra. Copper-Neon-Argon arcs were also
taken for the purpose of wavelength calibration.

For one candidate, SN~2002lk, first observed on June 10th, one more
exposure of 600s was taken on June 11th. Indeed, this very
unusual supernova at low redshift ($z=0.033$) was identified as a
spectroscopically underluminous supernova during a preliminary
reduction at the telescope (see below).

A summary of the spectroscopic observations is given in Table
\ref{tab1}.

\subsection{Spectra reduction and extraction}

Data reduction was done in two steps: first, a rapid assessment of the
spectra was performed at the telescope in order to decide whether
more telescope time should be spent on a particular object. For
example, after the "real time" analysis of SN~2002lk, we decided to
take another spectrum of this object during the next night.
Then, a full reduction of the spectra was done off-line, using
ESO-MIDAS data reduction software version SEP02. After debiasing and
flat-fielding, a two dimensional dispersion relation was obtained from
the identification of lines for each row of an CuNeAr arc frame (the
dispersion axis is horizontal in our settings) and applied to every
frame for wavelength calibration. Mild geometrical distortions along
the slit, likely due to instrument flexure during observations, were
traced and corrected for. The wavelength calibration was checked
against a few prominent emission sky lines, both for blue (NaI at
5893\AA$\ $ and [OI] at 5577\AA) and red ([OI] at 6300 and 6363\AA)
frames, and was found to be at most within 1$\AA\ $ of the tabulated
value. This lead to a wavelength calibration accuracy better than
0.05~\% over the entire wavelength scale. The wavelength calibrated
frames were then corrected from extinction using the extinction curve
of La Palma observatory \citep{King85} and absolute flux calibration
was done using spectrophotometric standards spectra acquired at
the beginning and end of each night. Sky subtraction was performed by
polynomial interpolation of the sky signal at the location of the
spectrum, using two windows above and below the candidate spectrum.
Parameters for sky extraction were optimized in order to minimize the
r.m.s. dispersion of the sky residuals in the sky subtracted frame, in
200$\AA\ $ wavelength bands centered on $4200\AA\ $ and $5900\AA\ $
for the blue arm, $6500 \AA\ $, $7100\AA\ $ and $8100 \AA\ $ for the
red arm, where sky emission lines are important.  Extraction of
spectra was then performed following the \citet{Horne86} optimized
algorithm. Sky residuals were subtracted by hand but we did not attempt
to correct for atmospheric absorption features. An error spectrum was
computed for each candidate spectrum using the CCDs noise and gain
properties. Error spectra are dominated by the sky emission both in
the red and blue spectra, but, due to the dichroic response, the error
is slightly larger in the blue than in the red.

\section{Identification procedure}

\subsection{Redshift determination}

As a first step in identification, we determined the redshift of
the candidate from galaxy emission or absorption lines if present, or
directly from the supernova absorption features. If derived from
galaxy lines, the accuracy on the wavelength calibration translates
into an absolute uncertainty on the redshift of less than 0.001 at
$z=0.5$ and even lower at intermediate redshift. As a consequence, all
redshifts derived from the host lines (denoted $z_{h}$) are given in
this paper with three significant figures. In the case where the
  redshift determination is drawn from supernova features, the
  uncertainty increases by up to 0.01 due to large P-Cygni lines and
  dispersion in the line velocities among supernovae at a given phase.

\subsection{Template fitting}

As for our 1999 campaign, we used a template fitting technique based on
a $\chi^2$ minimization to identify SN candidate spectra. Using ${\cal
  SN}$-fit, software developed for the SNLS collaboration
\citep{Sainton04a,Balland06}, a series of model templates is built as
a combination of a fraction of a supernova and a fraction of a galaxy
properly redshifted to the redshift of the spectrum candidate (if
known, otherwise the redshift is left as a free parameter):

\begin{equation}
\label{eq1}
{\cal M}(\lambda_{rest},z,\alpha,\beta)=\alpha{\cal S}(\lambda_{rest}(1+z))
+\beta{\cal G}(\lambda_{rest}(1+z)).
\end{equation}

These models are compared to the spectral data. The $\chi^2$
minimization is done against the pair of parameters $\alpha$ and
$\beta$, excluding undesirable regions, such as atmospheric
absorptions, if necessary, and is performed for all couples SN/galaxy
available among the different template categories selected by the
user. Solutions are ranked in order of increasing reduced $\chi^2$ and
the resulting parameter set and templates are given.  The stability of
the phase (and of the redshift obtained from the fit when no host
lines could be used) is checked by examining the five first solutions
immediately following the best-fit solutions. The typical error
  on the phase determination for \Ia spectra around maximum (up to
  $\approx 15$ days) was found to be $\pm 3$ days for spectra with
  comparable signal-to-noise ratio as in \citet{Balland06}. This
increases to $\pm 5$ days for later phase spectra as supernova spectra
evolve less rapidly at later time in the photospheric phase. SN and
galaxy templates are drawn from a database\footnote{Appendix A of
  \citet{Balland06} describes this database in details} containing
about 250 templates including local Branch-normal \Ia with a
continuous phase sampling between -10 up to +15 days, under and
overluminous SN~Ia, a series of partially synthetic templates with a
phase sampling of $1$ day \citep{Nobili03}, various templates of type
Ib, Ic and II supernovae, and a set of galaxy templates from
\citet{Kinney96} representing the Hubble sequence from ellipticals to
late-type spirals.

All template spectra were dereddened, except for the two highly
reddened supernovae SN~1986G \citep{Phillips87} and SN~1998bu
\citep{Jha99}, and put into the restframe. No reddening correction is
allowed during the fit because such a correction is likely to be hard
to interpret, as many effects can combine to give a spectrum "redder"
than real (e.g., error in flux calibration, flux losses due to
differential refraction unproperly accounted for). However, one might
be worried that a reddened spectrum leads to a wrong, later phase than
real, identification. To test this hypothesis, \citet{Balland06} have
artificially reddened by various amounts of $E(B-V)$ the template
spectrum SN~1994D \citep{Patat96} at -2 days and fitted the reddened spectra. They
found that, as higher values of $E(B-V)$ are used, the fit tends
indeed to produce later phase (redder) solutions. However, for values
higher than 0.2, no satisfactory fit is obtained. For $E(B-V)<0.2$,
they found that this reddening-phase degeneracy can not lead to an
error on the phase of more than 2 to 3 days, comparable with the accuracy
on the phase determination obtained for unreddened spectra.

 The fit is performed on the wavelength range over which the template
 model and the spectral data overlap. Even if the database templates
 have been selected, among other criteria, for their large spectral
 range, once redshifted, the overlap region with the data might be
 small. In that case, confusion with other types is possible as, e.g.,
 type II supernova features might fit well the red part of a SN~Ia.
 Another possible degeneracy that might falsify the identification is
 the possible confusion between a SN~Ic at early phase with a \Ia
 after maximum. Complementary information, such as the date of B-band
 maximum light, should be used if available.

\subsection{Host galaxy subtraction and identification}

Contamination of the supernova spectrum by its host signal often
occurs and it is an important step to subtract the galaxy contribution
in order to get the true supernova spectrum. In our scheme, the
contribution of the template galaxy to the total {\em model} is done
by ${\cal SN}$-fit. The fraction $\beta/(\alpha+\beta)$ represents the
fraction of the galaxy template in the model and not the real
contribution of the galaxy to the total signal. Nevertheless, the
fraction given by ${\cal SN}$-fit provides a useful indication of the
galaxy contribution. In case a spectrum of the host can be extracted
from the data, it can be used as an input galaxy to build the model.
However, this situation never occured in the present analysis.

Host identification is preferentially made from emission and/or
absorption lines when present in the spectrum. We adopt the division
of the host galaxy into three main morphological classes as proposed by
\citet{Sullivan03}: type 0 for spheroids (E/S0/bulge), type 1 for
early-type spirals (Sa/Sb) and type 2 for late-type spirals and
starbursts (Sc/Stb).  Spectral features used for the identification
(and for redshift determination; see \S 4.1) include Ca\,{\sc ii} H\&K
absorption lines at $3934$ and $3968$\AA, the 4000\AA$\ $ break
($B4000$), Hydrogen Balmer lines (mostly $H\beta$ and $H\gamma$ given
the redshift range of the present spectra), oxygen forbidden lines
[O\,{\sc ii}] and [O\,{\sc iii}]. When no line is present in the total
(SN+host) spectrum, identification is based on the result of the
minimization procedure. Indeed, ${\cal SN}$-fit selects a best-fit
host type according to the lines and the overall spectral energy
distribution.  This, in itself, gives an indication of the real nature
of the host and the concordance of the identification based on galaxy
lines with the best-fit galaxy from ${\cal SN}$-fit is found to be
satisfactory in approximately 80\% of occurences. We also determine a
host type from the SDSS colors available for our galaxies (see \S 5.4
below).

\subsection{Strategy for supernova identification}

In order to confirm the candidate nature, type and phase, we
  systematically perform a fit for the total (blue + red) spectra
  using all supernova and galaxy templates available. If a
  satisfactory solution is obtained for a \Ia template, we then
  perform three series of fits: one with Branch-normal \Ia templates
  alone, one with Nobili's templates alone, and one with
    peculiar (under and overluminous) \Ia templates. If the best-fit
  solution is not obtained for a \Ia template, we perform another fit
  restricting to SN~II or SN~Ib/c templates to compare the best-fit
  result with adjacent solutions with the same SN type. If galaxy
lines are absent, a plausible redshift range is determined from the
absorptions visible in the spectrum of the candidate supernova, and is
used as an initial guess for the minimization procedure. In all cases,
we redo the fits after having cut all galaxy lines present in the
data. This gives a new redshift fit, $z_{f}$. The uncertainty on
$z_{f}$ is typically of 0.01, that is 10 times higher than the
uncertainty on the redshift $z_h$ drawn from galaxy lines.

Comparison of the reduced $\chi^2$ leads to the candidate identification.
When the lowest $\chi^2$ is obtained for a normal SN~Ia, we test the
stability of the phase by 1) checking the five solutions immediately
following the best-fit solution in terms of increasing $\chi^2$; 2)
checking that the solution obtained from the fit with Nobili's
templates is consistent (within the typical uncertainty on the phase
$\pm 3$ days) with the normal \Ia solution; this usually happens,
except when the normal \Ia templates yield a solution fitting a too
small wavelength range of the data: in that case, due to its large
wavelength coverage, the Nobili's templates are likely to give a
better estimate of the phase; 3) that fitting the red and blue spectra
separately yields consistent solutions in terms of phase; 4) that the
best-fit solution for the full spectrum is consistent with the
solutions obtained when fitting the blue or red part alone. This
insures that the phase is securely determined within $\pm 3$ to 4
days.

\section{Spectroscopic results}

\subsection{General results}

Table~\ref{tab2} presents the results of the identification for the
eight supernovae observed during this campaign. For each supernova,
the best fit result is given for the total spectrum. Columns 2 and 3
give the best-fit supernova and galaxy template model. In column 3,
the percentage of galaxy signal in the best-fit model is indicated.
Columns 4 and 5 give the best-fit redshift from the fit $z_{f}$ and
from host lines $z_h$. Finally, columns 6 to 8 indicate the phase of
the best-fit solution, the reduced $\chi^2$ value and the associated
number of degrees of freedom. Five out of the eight supernovae
observed are found to be normal \Ia (SN~2002li, SN~2002lj, SN~2002lp,
SN~2002lq and SN~2002lr), while one is identified as a
spectroscopically underluminous \Ia (SN~2002lk) and two are type II
supernovae (SN~2002ln and SN~2002lo). For the \Ia of the sample, three
are pre- or close to maximum light whereas three have phases between
3 and 9 days after maximum. The two type II supernovae in our
sample are at late phases. As SN~II are less energetic events than
SN~Ia, only those at relatively low redshift ($z\lesim 0.15$) match the
selection cuts of the present search and have thus been selected as
potential \Ia candidates.

\begin{figure*}
\resizebox{16cm}{!}{\includegraphics{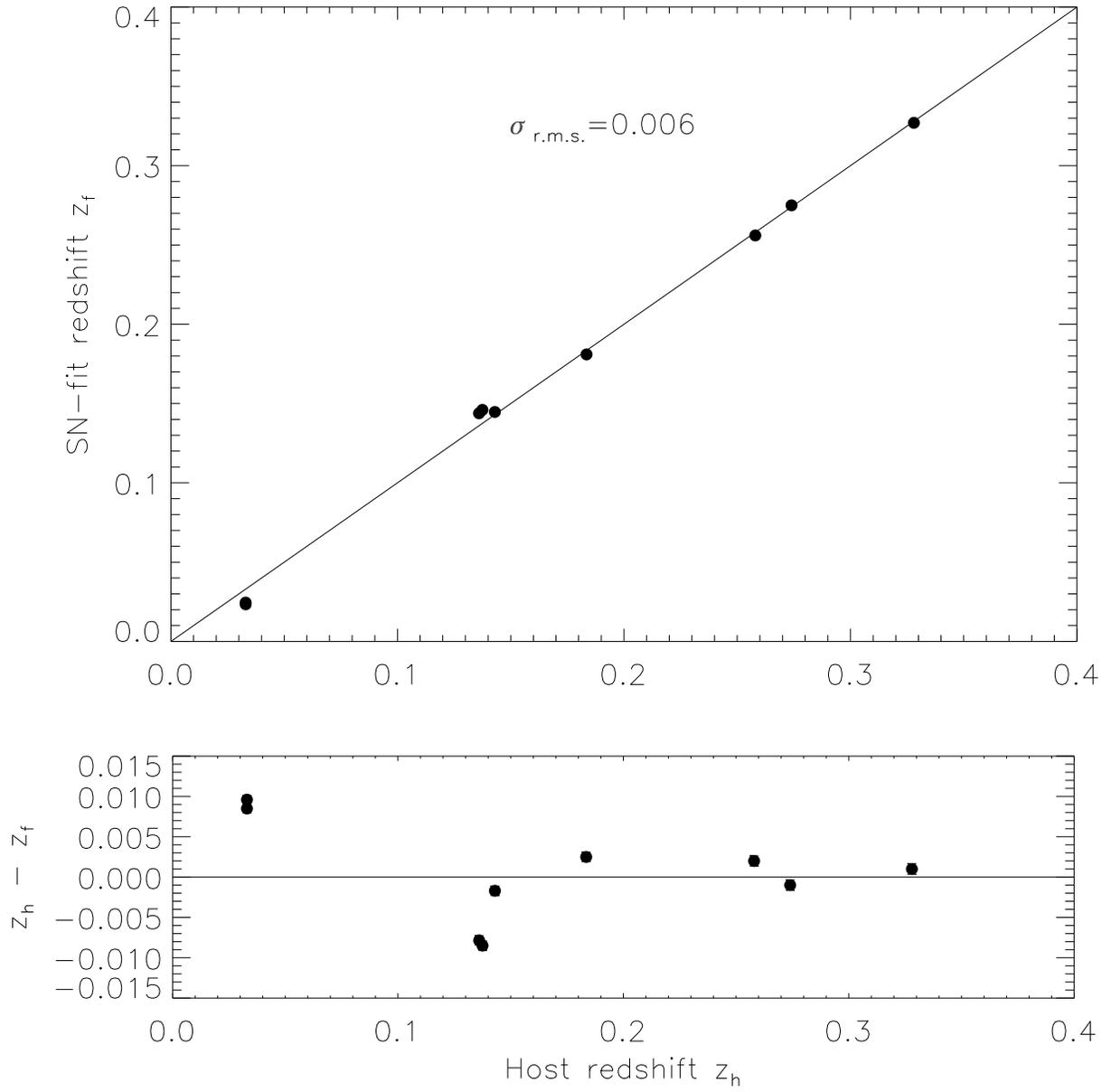}}
\caption{Redshift $z_f$ from ${\cal SN}$-fit as a function of the
  redshift $z_h$ of the host galaxy (top panel), and residuals
  (bottom panel). Error bars shown are for the host redshift only. The dispersion on $z_f$ is derived from this Figure.}
\vfill
\label{fig1}
\end{figure*}

In Figure~\ref{fig1}, we show the fitted redshift $z_{f}$ as a
function of the redshift $z_h$ (upper panel), and the corresponding
residuals (bottom panel). The two quantities agree at better than the
1\% level, with the dispersion being higher for supernovae with weak
galaxy features. The total r.m.s dispersion is $\sigma = 0.006$, and
is an estimate of the error on the redshift determined by ${\cal
  SN}$-fit. On the residuals, it is seen that no systematic effect
affects the redshift determination.  All observed supernovae are in
the range $0.033<z<0.33$, with an average redshift $<z>\sim 0.18$.
Excluding the peculiar SN~2002lk and the two type II-P SN~2002ln and
SN~2002lo, the average redshift becomes $<z>\sim 0.24$. The redshift
of SN~2002lk is found to be $z_h=0.033$. Indeed, the selection of an
underluminous supernova implies it to be at a much lower redshift than a
normal SN~Ia. SN~2002lk is thus not in the intermediate redshift range
targeted during this search and we will exclude it from the
intermediate $z$ sample analysis. However, as a peculiar object, it is
an interesting supernova and we devote to it a specific analysis
later in this paper.

\begin{figure*}[t]
\resizebox{16cm}{!}{\includegraphics{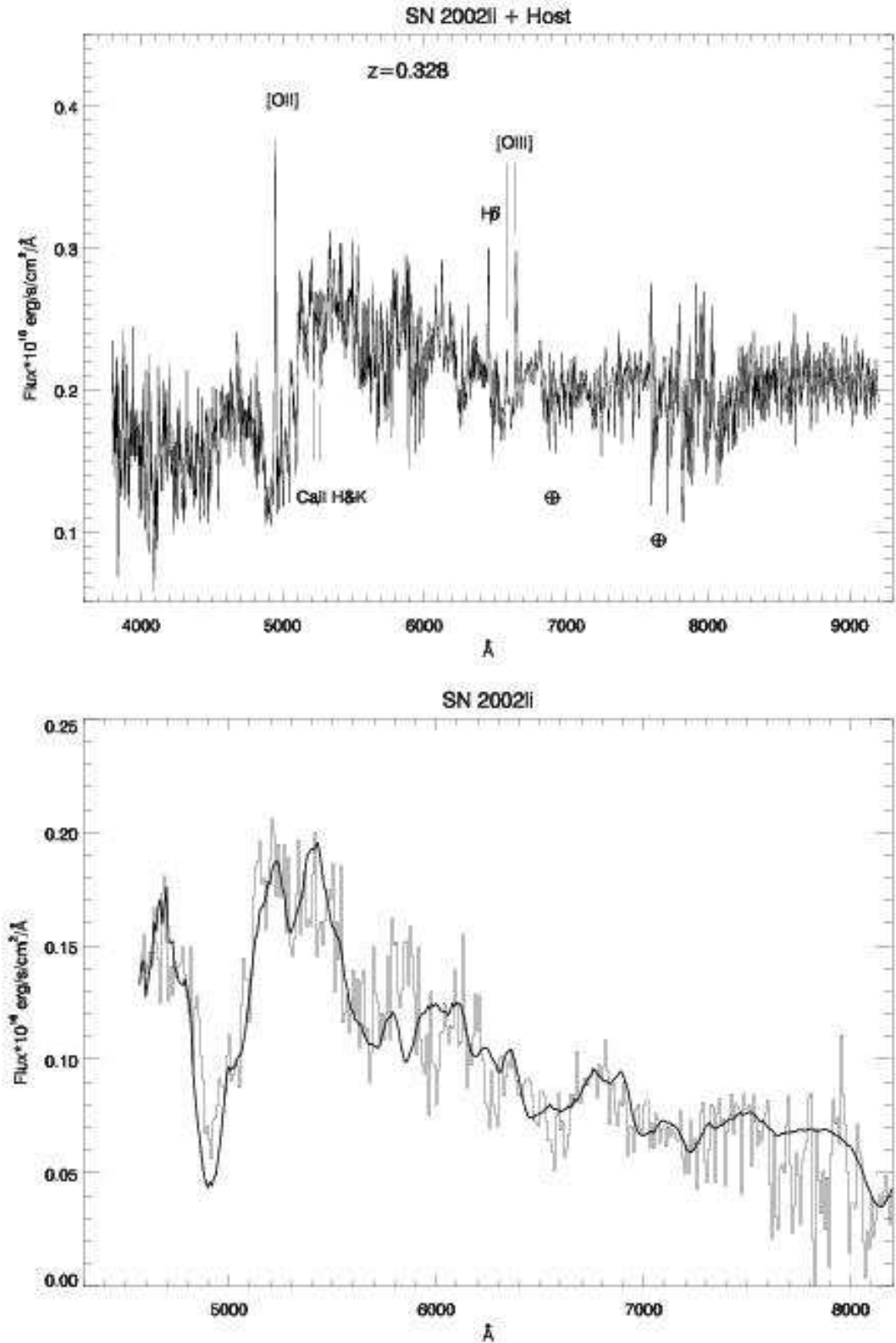} }
\caption{Spectrum of SN~2002li in the observer frame. Top: Full host+SN spectrum. The host redshift is indicated. Bottom: SN spectrum with the best-fit normal Ia template (SN~2003du $-$7
  days) overlapped.
   The SN spectrum has been rebinned to 10\AA$\ $ bin for visual convenience. It is corrected for galactic line subtraction residuals, but not for atmospheric absorptions.}
\label{fig2}
\end{figure*}

\begin{figure*}
 \resizebox{16cm}{!}{\includegraphics{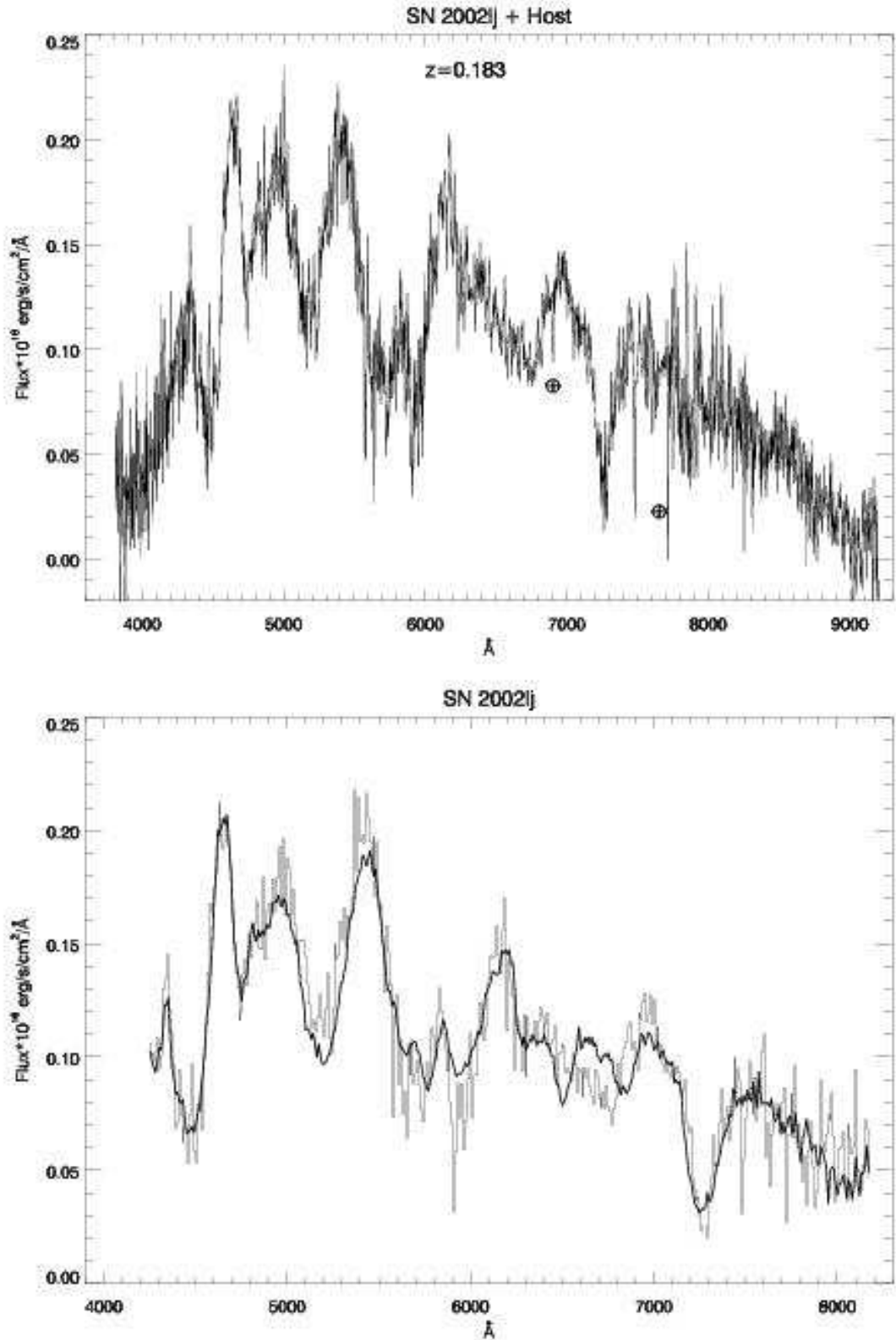} }
\caption{Same as Fig. \ref{fig2} for SN~2002lj. The best-fit template for the full spectrum is SN~1992A +7 days}
\label{fig3}
\end{figure*}

\begin{figure*}
 \resizebox{16cm}{!}{\includegraphics{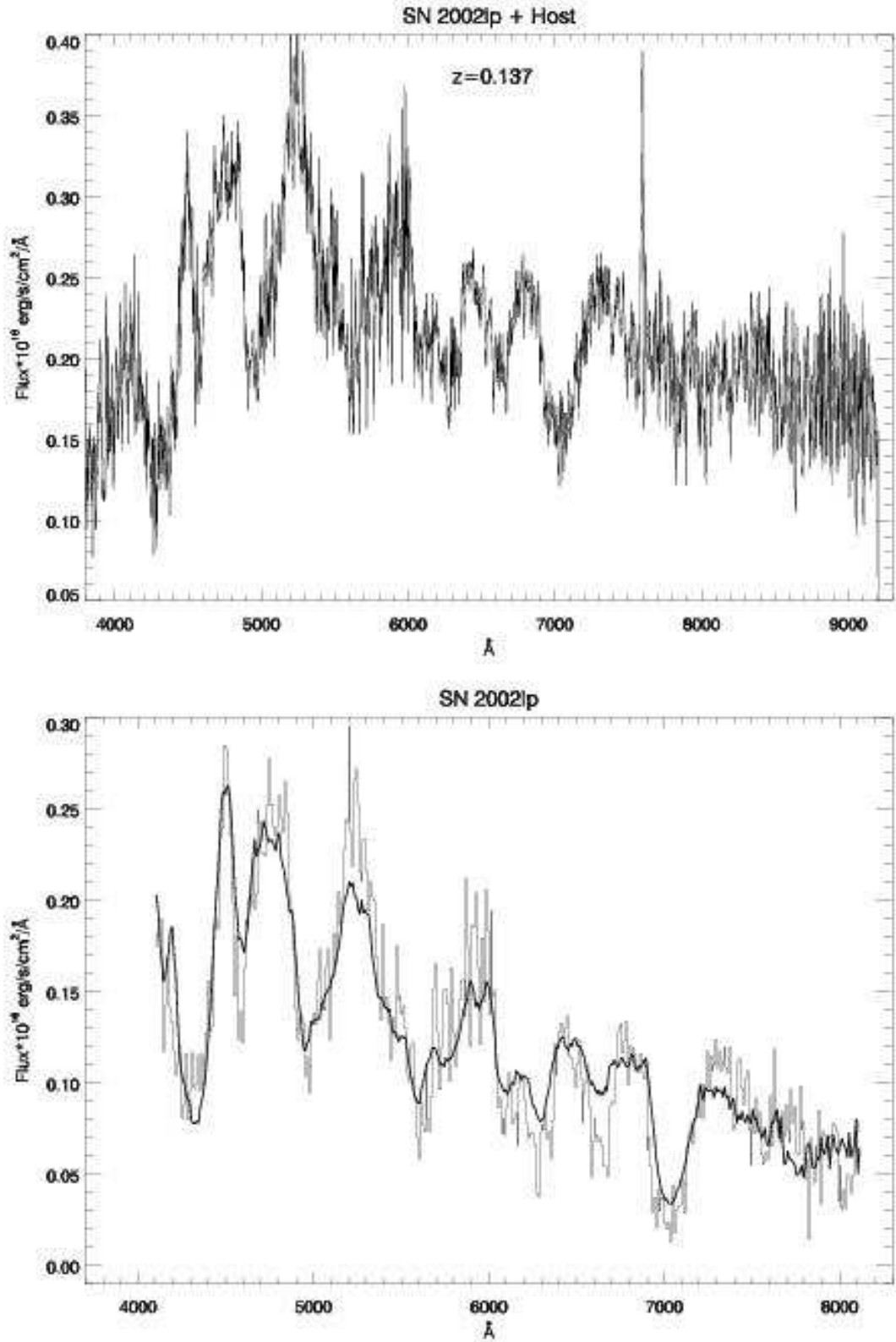} }
\caption{Same as Fig. \ref{fig2} for SN~2002lp. The best-fit template is SN~1992A +3 days.}
\label{fig4}
\end{figure*}

\begin{figure*}
\resizebox{16cm}{!}{\includegraphics{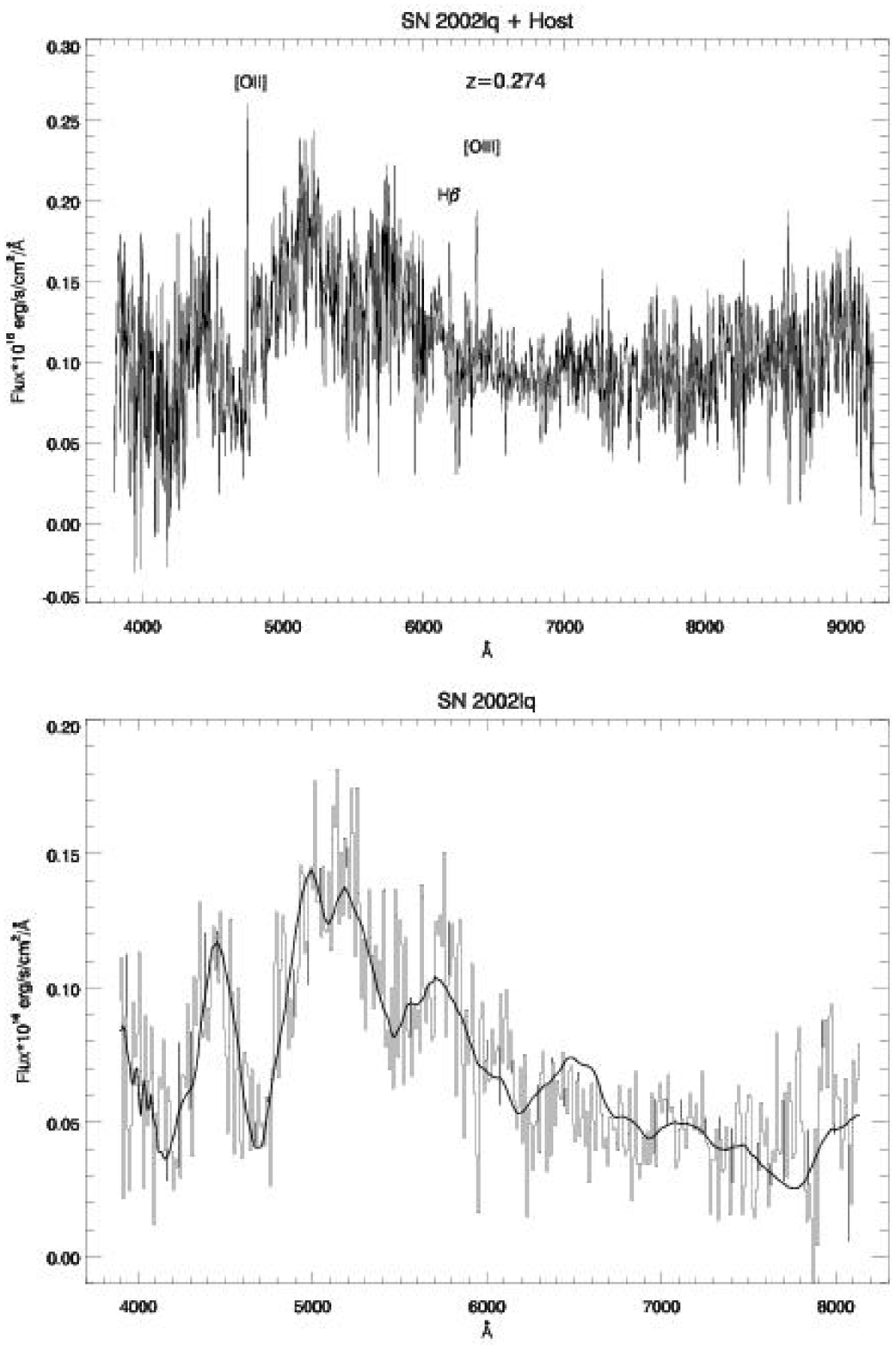} }
\caption{Same as Fig. \ref{fig2} for SN~2002lq. The best-fit template is SN~1994D $-$11 days. Here, a Nobili  template at $-$11 days, covering a larger spectral range, is shown.}
\label{fig5}
\end{figure*}

\begin{figure*}
  \resizebox{16cm}{!}{\includegraphics{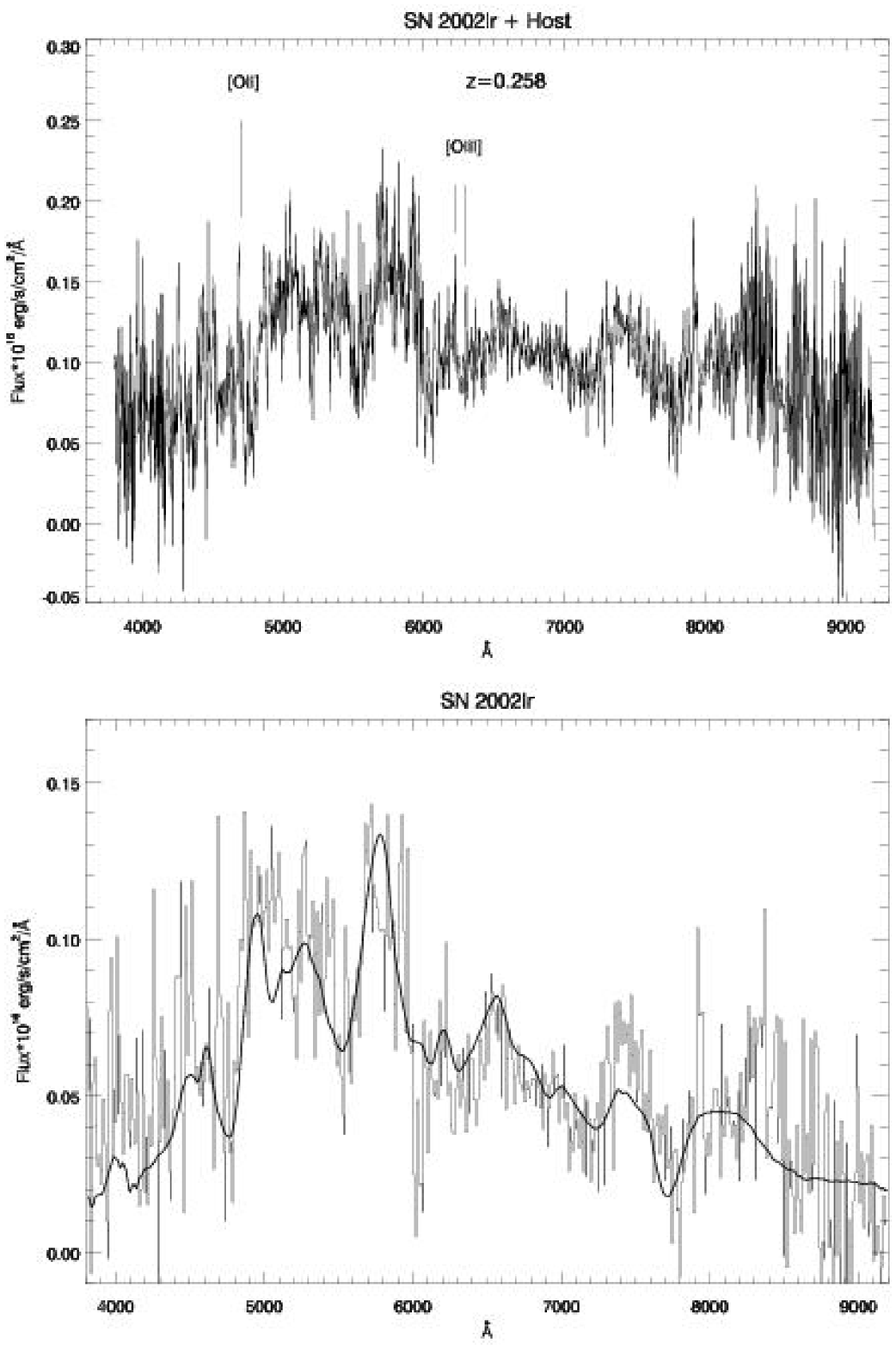} }
\caption{Same as Fig. \ref{fig2} for SN~2002lr. The best-fit template is SN~1992A +9 days. Here, a Nobili template at +9 days, covering a larger spectral range, is shown.}
\label{fig6}
\end{figure*}

In Figures \ref{fig2} to \ref{fig10}, we present the spectra. For each
supernova, we show the total spectrum (SN+host) over the full spectral
range (top panel), and the SN alone with the best-fit solution
superimposed (bottom panel). All spectra are presented in the observer
frame and the SN spectra have been rebinned on 10\AA$\ $ bins for
visual convenience. Uncorrected atmospheric lines are indicated with
the sign $\oplus$. When visible, galactic lines are labeled with the
corresponding ion.  A smoothed version of these \Ia spectra is
presented in Figures \ref{fig11} and \ref{fig12}, for pre-maximum and
post-maximum supernovae respectively. The spectra are ordered with
increasing phase. Residual lines resulting from imperfect galaxy
subtraction and atmospheric absorption lines have been removed, and
the spectra have been smoothed using a Savitsky-Golay filter of degree
2 with a 60 data points window \citep{Press86}. The three grey lines
on these two latter graphs indicate the Ca\,{\sc ii}, S\,{\sc ii} and
Si\,{\sc ii} found in normal SN~Ia. In addition, three vertical solid
lines indicate the positions of Ca\,{\sc ii} at 3945\AA, S\,{\sc ii}
at 5640\AA$\ $ and Si\,{\sc ii} at 6355\AA, blueshifted by 15000
km/s (Ca\,{\sc ii}) and 10000 km/s (S\,{\sc ii}, Si\,{\sc ii}).  These
values are typical of 'normal' \Ia at maximum \citep{Benetti04} and
serve as a reference for visual inspection.

The presence of features due to intermediate mass elements in the
supernova spectra is a discriminant for the fits, and thus for the
identification. Due to the redshift range and spectral response of the
grating used with the red arm, the Si\,{\sc ii} signature found at
6150\AA$\ $ (in the restframe of the supernova) is often noisy if
visible at all.  The W-shape around 5500\AA$\ $ due to S\,{\sc ii}
absorption at 5640\AA$\ $ often falls close to or in the cut-off
region of the dichroic filter. Moreover, this feature is observed
around maximum and fades at later epochs. It is indeed often difficult
to identify it in our spectra. Other discriminant features in the blue
part of the spectra are the Ca\,{\sc ii} absorption at 3945\AA$\ $ and
the Si\,{\sc ii} at 4130\AA$\ $. Regarding SN~II, the main
discriminant features are the P-Cygni profiles of the Hydrogen Balmer
series, mostly $H\alpha$ and $H\beta$.


\subsection{Notes on individual \Ia}

In this section, we detail the properties of each spectrum and discuss
the results of the fitting procedure.

\subsubsection{Normal \Ia}

{\bf SN~2002li:} This is a typical \Ia before maximum (Fig.~\ref{fig2}). 
It has the largest redshift of the sample
($z_h=0.328$). The full spectrum is well fitted by SN~2003du
\citep{Anupama05} at $-$7 days. The phase is very stable when one goes
to higher rank solutions. The red part does not show many wiggles and
the fit is mostly sensitive to the overall energy distribution. The
best-fit solution for the red spectrum only is for SN~1999ee
\citep{Hamuy02} at $-$9 days. The same solution is found when the blue
spectrum alone is fitted by Nobili's templates. Again, the phase is
very stable with higher ranked solutions.  However, fitting the blue
spectrum with \Ia templates yields a somewhat lower value (SN~1999ee
at $-$4 days), but the fitting range is small (269 pixels) and
degeneracy between solutions may occur. We thus favor the $-$7 day
solution, however note that the solution at $-$4 days is only
marginally discrepant, given the $\pm 3$ day uncertainty.

The host galaxy contamination is fairly high as it ranges from $\sim
50$ to $\sim 65$\% for the three fits (blue, red, total spectrum).
The best-fit host type is a Sa galaxy in all cases. Despite a somewhat
noisy spectrum (high $z$ and fairly high host contribution), the
identification is quite secure given the observed stability on the
phase.
 
{\bf SN~2002lj:} The signal-to-noise ratio is excellent (13 over the
whole spectral range) for this typical \Ia a week after maximum light
(Fig.~\ref{fig3}). The redshift ($z_h=0.183$) is determined from
very weak host emission and absorption lines (the host contribution is
low --- less than 15\% of the galaxy template is needed to build the
best-fit model). The full spectrum is well fitted with SN~1992A, 7
days past maximum \citep{Kirshner93}. The phase is very stable as the
contiguous solutions to the best-fit are all for \Ia at +7 days. The
fit is excellent visually, except in the range 6100-6700\AA$\ $ where
the redshifted W-shape S\,{\sc ii} feature seen in \Ia spectra up to
one week after maximum is absent. This suggests that the phase of +7
days is a lower limit on the true phase. The fit of the sole red part
of the spectrum corroborates this fact, as the best-fit solution is
obtained for SN~1994D 10 days past maximum. Here again, this phase is
very stable, and fitting with Nobili's templates gives the same result
for the phase. The best-fit solution for the blue spectrum is for
SN~1992A +7 days and the best-fit phase for a fit with Nobili's
template is +8 days. The phase for the blue fit is very stable. All
the fits we have attempted on this supernova yield a phase between 7
days up to 10 days, two values consistent within our adopted 3 day
error bar.

The $\chi^2$ value obtained for the various fits is rather poor, a
fact that traces back to the good signal-to-noise ratio of this
spectrum.  Indeed, when the signal is good (the error is low), the
discrepancies between the model built by the fitting procedure and the
data yield a high value of the $\chi^2$. In some sense, the higher the
signal-to-noise ratio, the poorer the fit in terms of $\chi^2$ values.
This fact, already noted in \citet{Balland06}, underlines the limits
of reproducing a large variety of supernova spectra from the template
database.

As the galaxy contamination is weak, it is hard to derive any
significant conclusion regarding its type from the fit. We note
however that the best-fit is consistently obtained for a spiral
galaxy, either a Sa (blue and full spectrum fit) or a Sb (red fit).

{\bf SN~2002lp:} This supernova at $z_h=0.137$ is well fitted by
SN~1992A template at +3 days (Fig. \ref{fig4}). Here again, the
high $\chi^2$ value obtained reflects the discrepancies between the
spectrum and the model, exacerbated by the good signal-to-noise ratio
of the spectrum. Although the main features are convincingly
reproduced by the model, we note a significant discrepancy around the
Si\,{\sc ii} 5972$\AA\ $ as the model does not reproduce the deep
silicon absorption. This is similar to what is observed in
underluminous SN~Ia. However, in underluminous SN~Ia, this deep
Si\,{\sc ii} absorption goes along with another absorption at $4100$\AA$\ $ 
clearly not observed in our spectrum. We conclude that
SN~2002lp is a normal SN~Ia. The observed discrepancy illustrates the
large diversity found in \Ia.

{\bf SN~2002lq:} The features of this SN at $z_h=0.274$ are most
noticeable in the blue part of the spectrum with a prominent Ca\,{\sc
  ii} 3965 \AA$\ $ absorption and are consistent with a normal \Ia
about 10 days before maximum light (Fig. \ref{fig5}). The
best-fit phase is $-$11 days for SN~1994D. Features in the red part of
the spectrum are less prominent but the overall energy distribution is
consistent with this solution. Galaxy emissions are clearly detected
in this moderately host contaminated spectrum.

{\bf SN~2002lr:} The spectrum of this far ($z_h=0.258$) supernova is
rather noisy but clear \Ia features are visible both in the blue and
the red part of the spectrum (Fig. \ref{fig6}). The best fit
model (SN~1992A +9 days) reproduces fairly well the main features.
This best-fit phase is to be taken as a lower limit, the uncertainty
on the phase being larger from a week past maximum, as supernova
features evolve less. The fit is poor from 8300 to 8600$\AA$ (observer
frame). This is likely to be due to sky subtraction residuals rather
than to a real difference in the supernova spectrum, as sky emissions
start being very important at these wavelengths.

\subsubsection{A spectroscopically SN~1986G-like \Ia ?}

\begin{figure*}
 \resizebox{16cm}{!}{\includegraphics{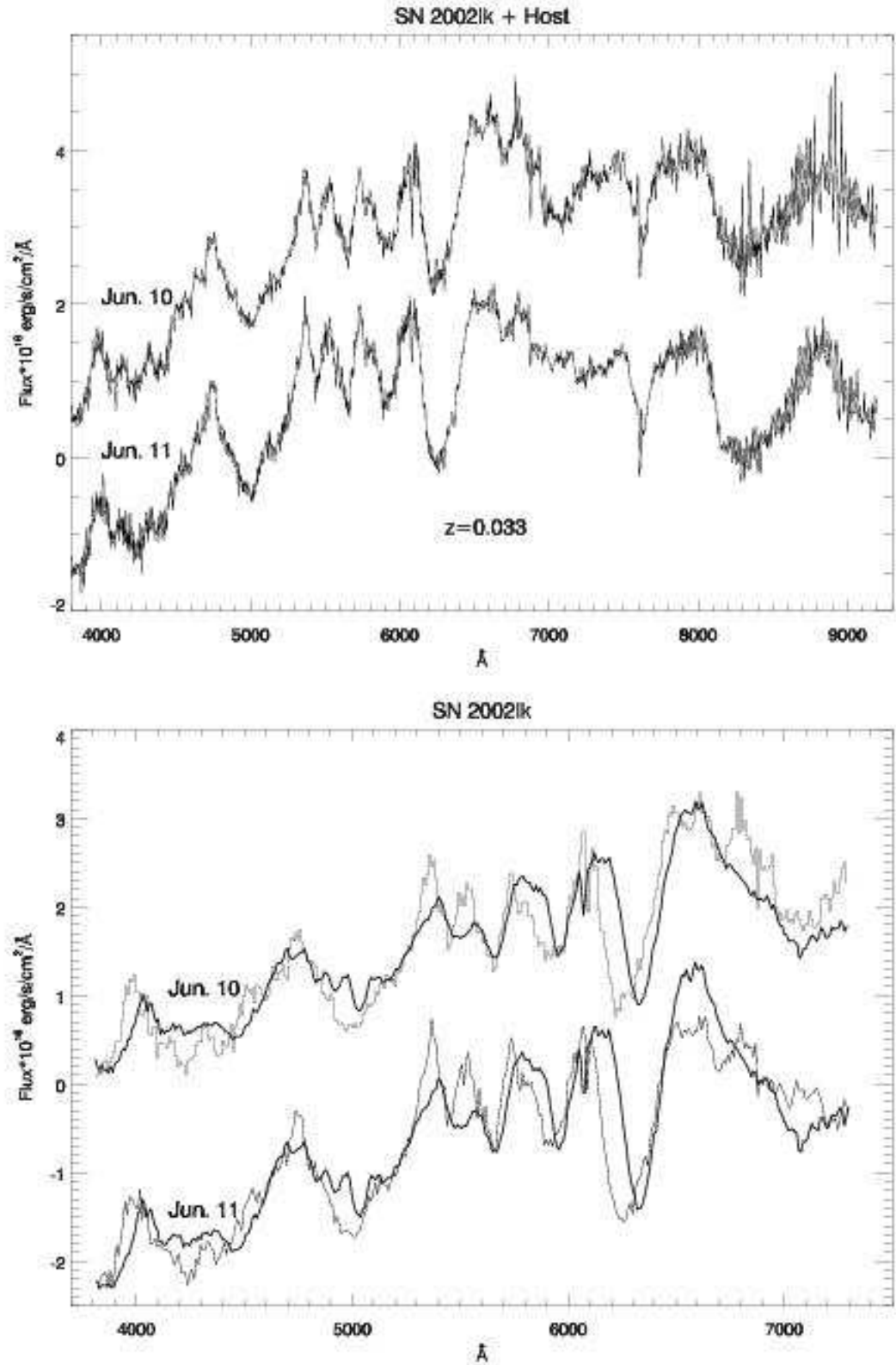} }
\caption{Same as Fig. \ref{fig2} for the June 10th and June11, 2002 spectra of SN~2002lk.  The templates overlapped to the spectra in the bottom panel are SN~1986G +2 days.}
\label{fig7}
\end{figure*}

\begin{figure*}
 \resizebox{16cm}{!}{\includegraphics{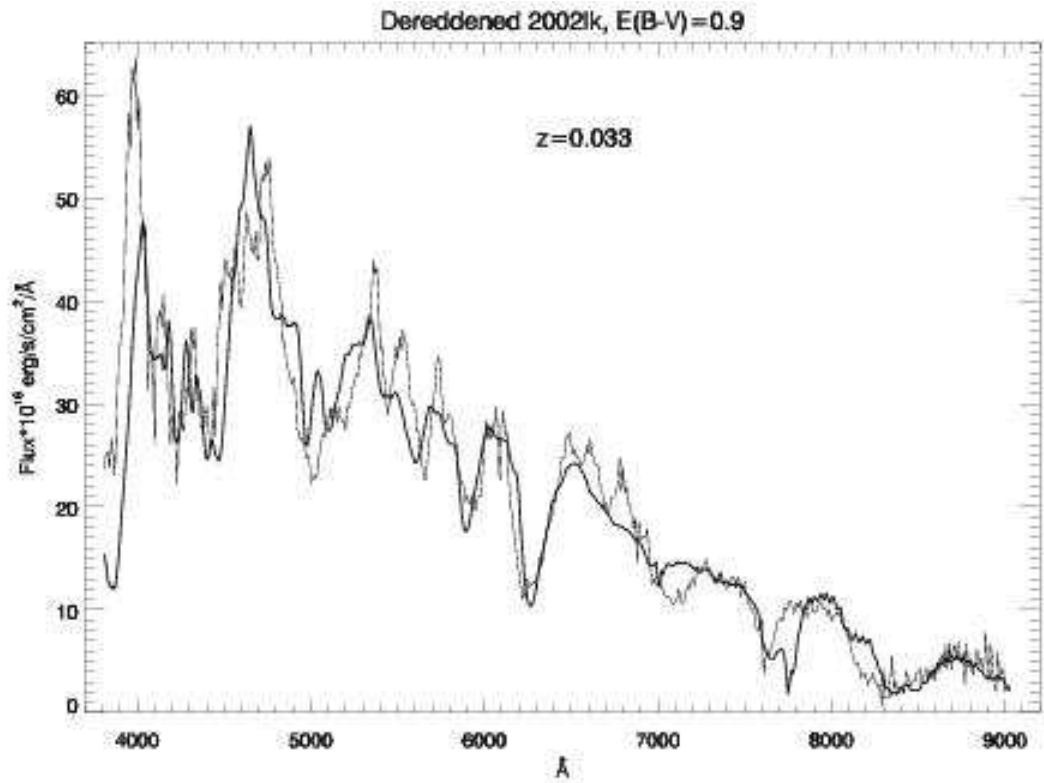} }
\caption{The June 10th 2002 spectrum of SN~2002lk has been dereddened using $E(B-V)=0.9$. The best-fit template is SN~1999by at $-$4 days.}
\label{fig8}
\end{figure*}

\begin{figure*}
 \resizebox{16cm}{!}{\includegraphics{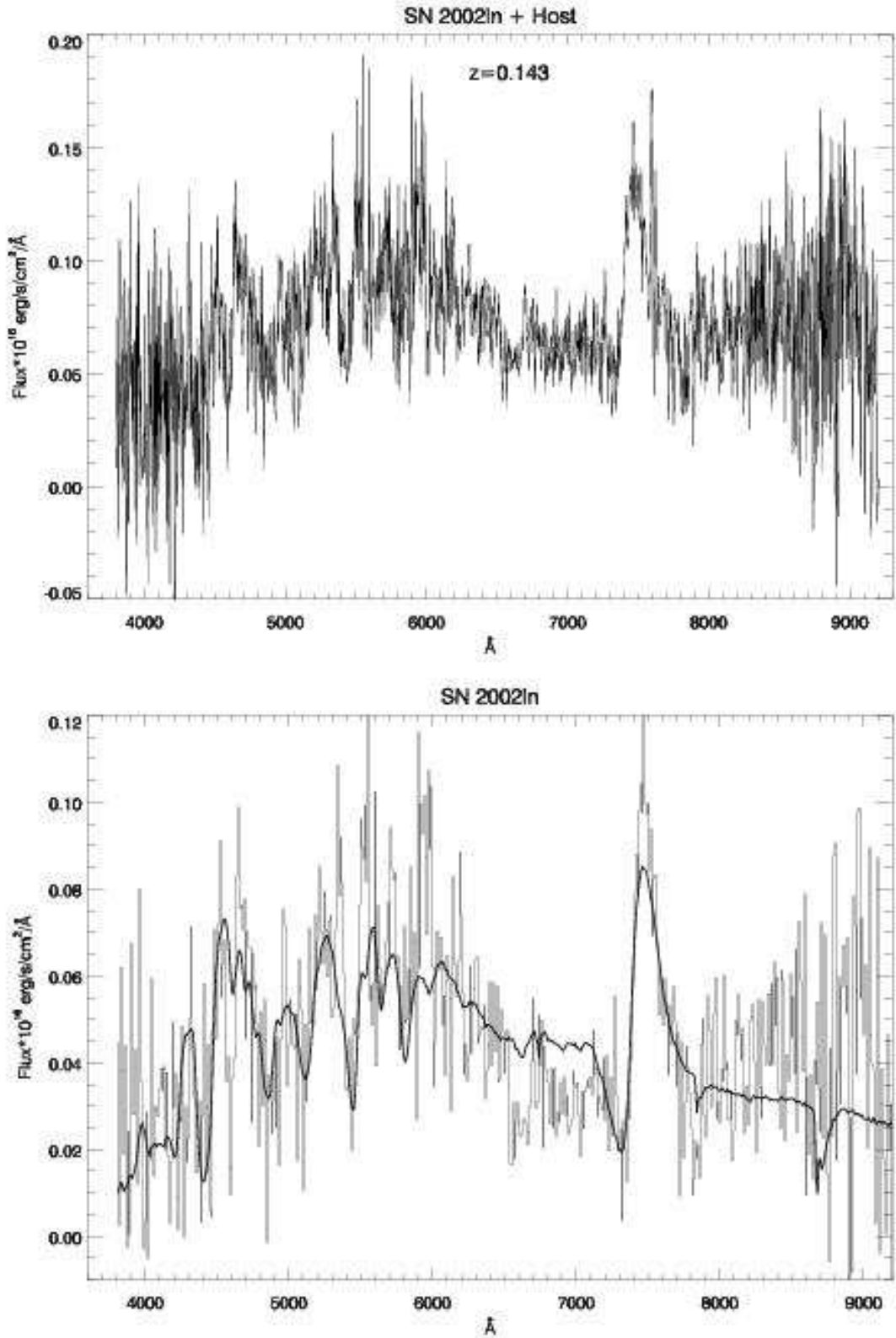} }
\caption{Same as Fig. \ref{fig2} for SN~2002ln.  The best-fit template is SN~1999em at +23 days after explosion.}
\label{fig9}
\end{figure*}

\begin{figure*}
\resizebox{16cm}{!}{\includegraphics{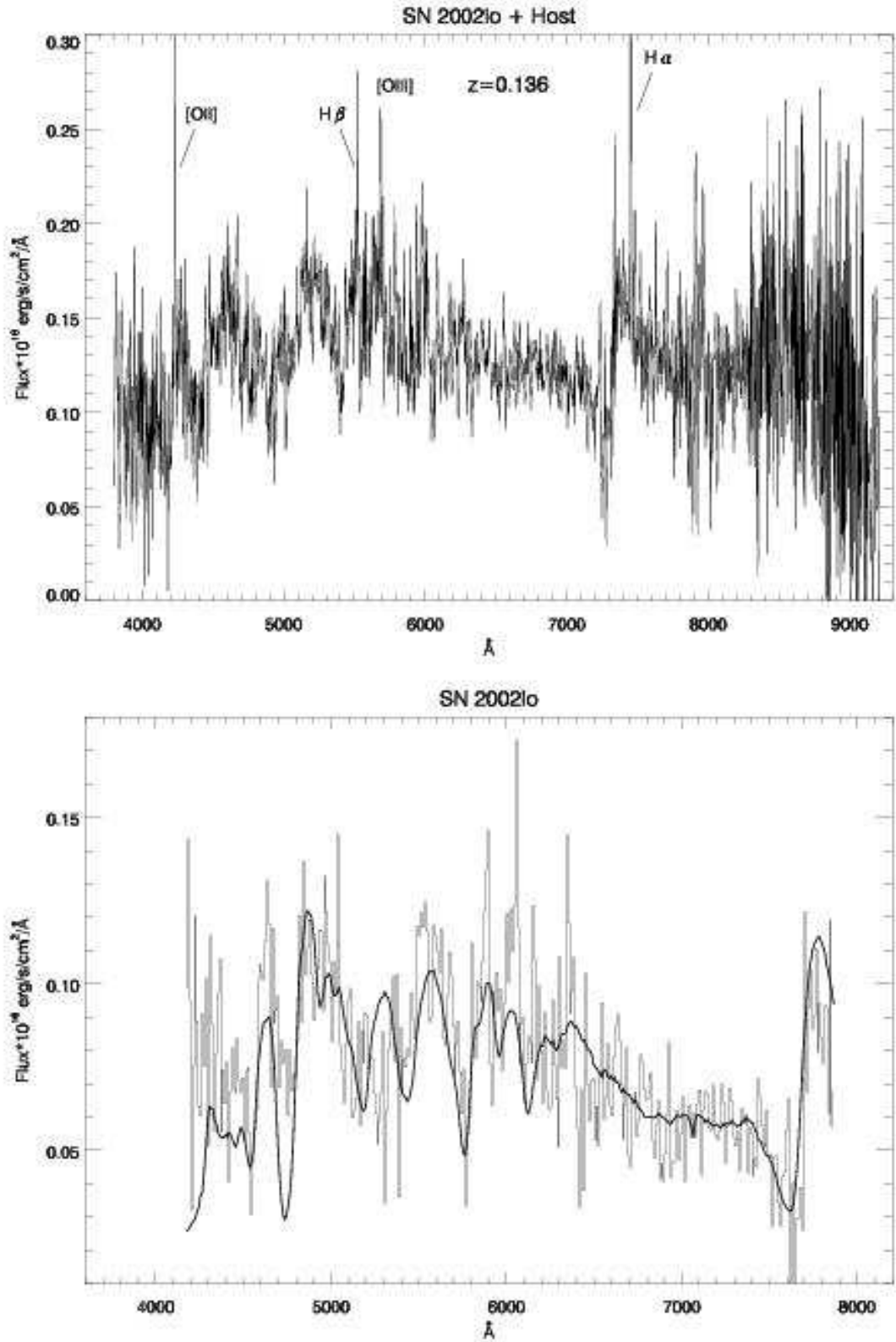} }
\caption{Same as Fig. \ref{fig2} for SN~2002lo. The best-fit template for the blue spectrum is SN~1999em +13 days after explosion.}
\label{fig10}
\end{figure*}

\begin{figure*}
\resizebox{16cm}{!}{\includegraphics{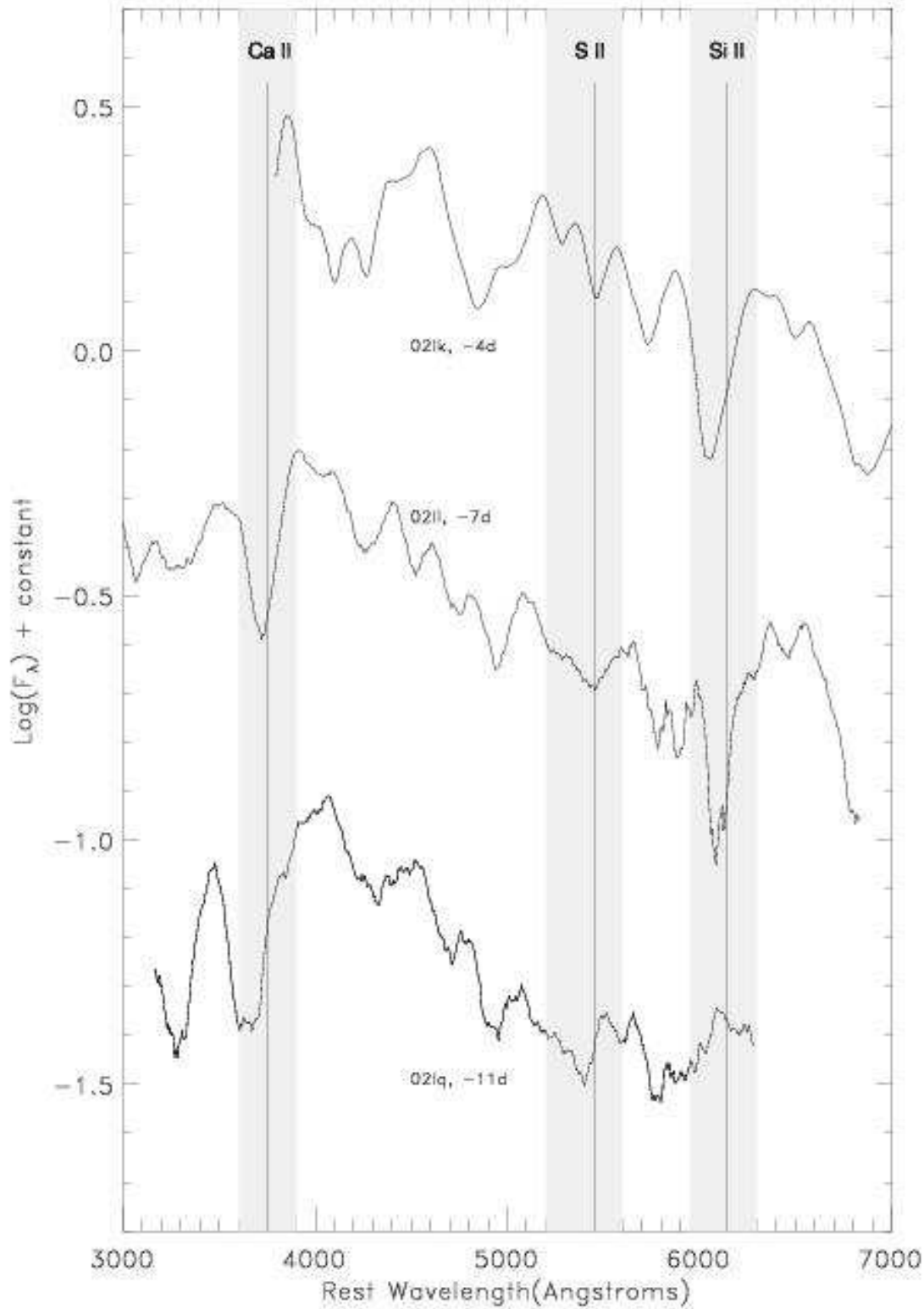} }
\caption{Heavily smoothed spectra of the three
  pre-maximum \Ia of the intermediate redshift sample ordered in a
  sequence of increasing phase (logarithmic scale). Spectra are
  shifted by an arbitrary amount for visual convenience, and are shown
  into the restframe.  For SN~2002lk, the deredddened spectrum of June 10th is shown. Atmospheric absorptions and galaxy line
  subtractions have been removed before smoothing.  Grey vertical
  bands show the Ca\,{\sc ii}, S\,{\sc ii} and Si\,{\sc ii} features found in normal SN~Ia.
  Solid vertical lines show the positions of Ca\,{\sc ii} at 3945\AA, S\,{\sc ii} at
  5640\AA$\ $ and Si\,{\sc ii} at 6355\AA$\ $, blueshifted by 15000 km/s
  (Ca\,{\sc ii}) and 10000 km/s (S\,{\sc ii}, Si\,{\sc ii}).  These values are typical of
  'normal' \Ia at maximum and are shown as a guide to the eye.}
\label{fig11}
\end{figure*}

\begin{figure*}
 \resizebox{16cm}{!}{\includegraphics{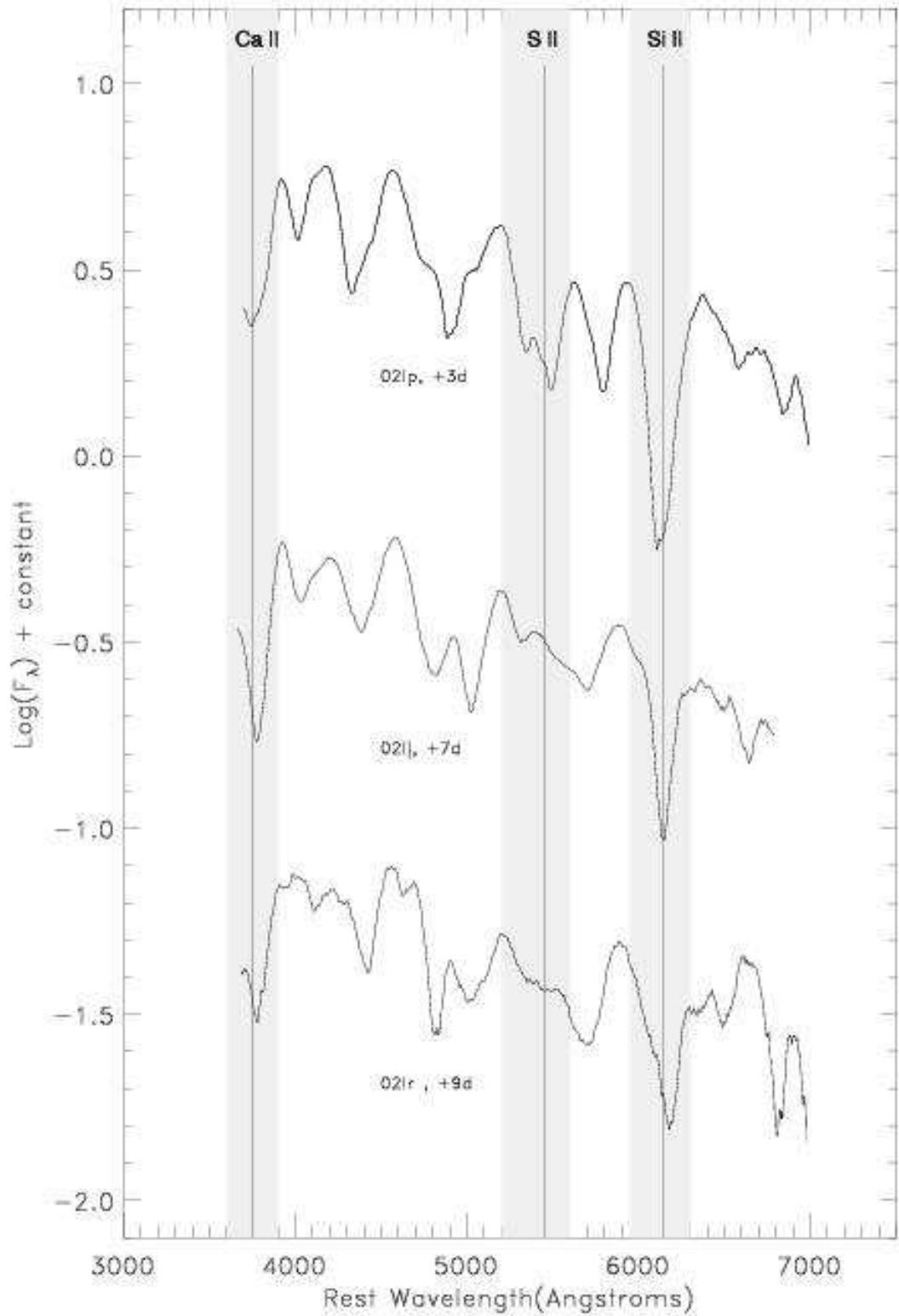} }
\caption{ Same as Fig. 11 for the three past-maximum 
  \Ia of the intermediate redshift sample.}
\label{fig12}
\end{figure*}

{\bf SN~2002lk:} The spectrum of SN~2002lk exhibits the features of a
reddened underluminous \Ia (Fig. \ref{fig7} and Fig. 
  \ref{fig8}). In many respects it resembles the spectrum of SN~1986G
\citep{Phillips87} as it lies in the dust lane of an early
spiral galaxy (probably Sb) visible on the discovery images of this
supernova. The spectral similarity is clearly visible in Figure
\ref{fig7}. Best fits of these reddened spectra are obtained for a
phase slightly past maximum (+2 days). Velocity shifts between the
spectra and the template are noticeable, with SN~1986G having smaller
blueshifts than SN~2002lk except around the sulfur W absorptions for
which the expansion velocities are comparable. The observed
discrepancies suggest different explosion energetics for these two
supernovae. Also, we note that the \ion{Si}{ii} and the
  \ion{Ca}{ii} IR triplet absorptions are very large as compared to
  the same features in the SN~1986G template.

In Figure \ref{fig8}, we show the best-fit obtained for the June 10th
spectrum of SN~2002lk dereddened using the reddening law of Howarth
(1983) and $E(B-V)=0.9$ (see \S 5.5 for more details). With this high
color excess, a fair match of the spectrum is obtained for SN~1999by
\citep{Garnavich04} 4 days before maximum.  Note the difference
between the phases obtained with fitting the reddened and the
dereddened spectra. We adopt the phase of this latter, which is in
almost perfect agreement with our preliminary photometric estimate of
the date of maximum light (see \S 5.3.3 below).

\subsubsection{SN~II}

As byproducts of our Ia supernova search, two candidates turn out to
be type II supernova.

{\bf SN~2002ln:} This noisy spectrum at $z_h=0.143$ is characterized
by the prominent $H\alpha$ P-Cygni profile found in type II SNe
spectra (Fig. \ref{fig9}). The best-fit is obtained for
SN~1999em, a type-II plateau supernova \citep{Hamuy01,Elmhamdi03}. The
best-phase is for about three weeks (23 days) after explosion.
However, this value is rather uncertain due to the scarcity of the
database templates at these 'late' phases. Besides $H\alpha$, we are
able to fit several features in the blue part of the spectrum despite
the fairly poor signal-to-noise ratio.

{\bf SN~2002lo:} The spectrum of SN~2002lo exhibits a $H\alpha$
P-Cygni profile less prominent than for SN~2002ln, but has blue
features more visible (Fig. \ref{fig10}). This type II-P
supernova at $z_h=0.136$ is best-fitted by SN~1999em about 10 days
after explosion (best-fit phase +13 days) although again some
uncertainty on this value remains.
 

\subsection{Properties of the SN Ia}

In this section, we analyse the spectra presented in this paper and
characterize their properties. Our main purpose is to assess the
homogenity of this new sample at intermediate redshift for comparison
to other supernovae at all redshifts.

\subsubsection{Velocity measurements}

\begin{figure*}
 \resizebox{16cm}{!}{\includegraphics{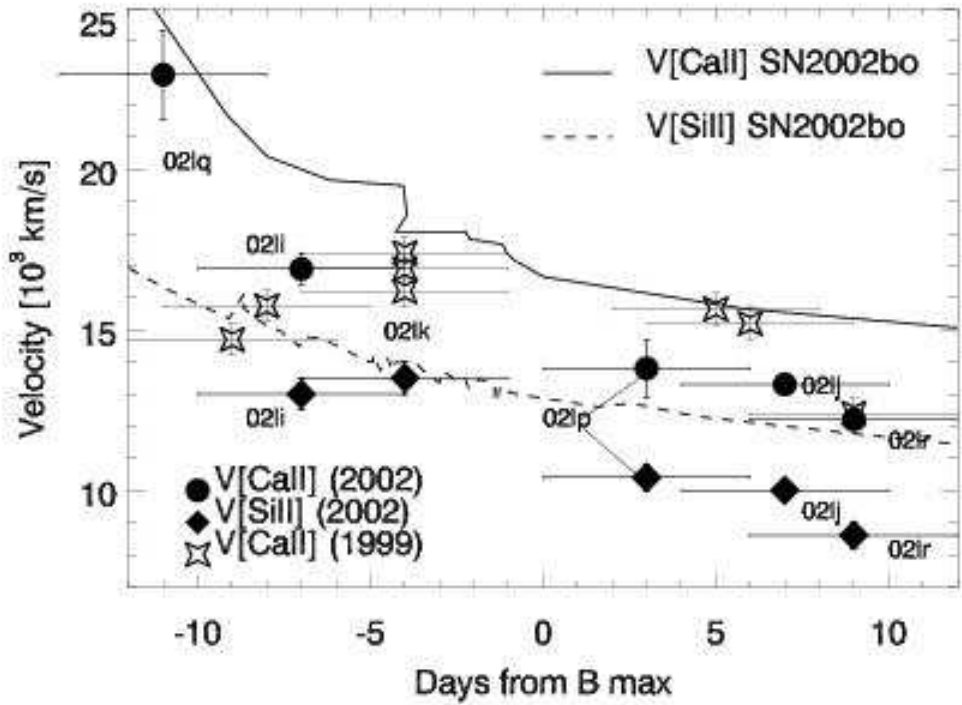} }
\caption{Ca\,{\sc ii}  and Si\,{\sc ii} velocities as a function of B-band phase.
  Solid and dashed lines are for SN~2002bo Ca and Si velocities respectively (data from Figure~11 of Benetti \etal (2004)). Calcium (filled circles) and Silicon (filled diamonds)
  velocities for the \Ia presented in this paper are given in Table 3. Also shown are the Calcium velocity measurements for the 1999 run supernovae \citep{Balland06}.}
\label{fig13}
\end{figure*}

Supernova expansion velocity measurements at a given epoch, or the
gradient between two different epochs, appear to be one of the major
indicators of the diversity of \Ia samples
\citep{Patat96,Benetti04,Benetti05,Hachinger06}.  We have been able to
measure a number of line velocities in our spectra for the calcium and
silicon features. Results are given in Table \ref{tab3} and reported
in Figure \ref{fig13}.  Errors on the velocity measurements are obtained
from several measurements of the minimum of the given absorption line
and summed quadratically. For the error on the date of maximum light,
we adopt the typical $\pm3$ days from the spectroscopic
identification. Solid and dashed lines are for SN~2002bo Ca\,{\sc ii}
and Si\,{\sc ii} respectively \citep{Benetti04} and are shown for
reference. The measurements made on the 2002 spectra are plotted as
filled circles (Ca\,{\sc ii}) and diamonds (Si\,{\sc ii}). Also
plotted as empty stars are the calcium measurements for 1999 \Ia taken
from \citet{Balland06}. The velocities of \ion{Si}{ii} for SN~2002lj,
SN~2002lp and SN~2002lr are similar to the one of SN~1994D
\citep{Patat96}. The \ion{Si}{ii} velocity of SN~2002li is slightly
higher. Only SN~2002lk exhibits a \ion{Si}{ii} velocity comparable
to SN~2002bo. The \ion{Si}{ii} velocity of SN~2002lk is $\sim
14000\mathrm{km}\mathrm{s}^{-1}$, about 3000 $\mathrm{km}
\mathrm{s}^{-1}$ higher than for SN~1986G. Regarding \ion{Ca}{ii}
velocities, all measurements are consistent with SN~1994D, with
SN~2002lq velocity being slightly higher. Measurement of \ion{Ca}{ii}
velocity for SN~2002lk is hardly possible, as this feature is located
at the lower limit of the effective spectral range due to the very low
redshift of this supernova. Analysis of line velocities does not
provide any hint of any peculiarity for the 2002 supernovae in the
range $0.15\lesim z \lesim 0.35$.  Even the fairly high \ion{Si}{ii}
blueshift measured for the spectroscopically subluminous SN~2002lk
falls within the dispersion observed among supernovae classified as
'Branch-normal' \Ia.

\subsubsection{Searching for peculiar SN Ia}

Putting aside SN~2002lk, we have systematically searched for
peculiarities in the spectra of the 2002 intermediate redshift sample
in order to assess the existence of peculiar \Ia at these
redshifts.  We follow the analysis performed in \citet{Balland06} and
systematically perform a fit using only the peculiar (under- or
overluminous) templates of our database. The best-fit solutions are
compared to the best-fits obtained using only normal \Ia templates.
The possibility that two solutions are equivalent in terms of $\chi^2$
is assessed by performing a F-test on the two solutions. The results
are presented in Table \ref{tab4} for the five \Ia of our sample in
the range $0.15\lesim z \lesim 0.35$. Columns 2 to 5 are for the fit
using peculiar templates, columns 6 to 8 are the best-fit results
using normal templates, as presented in Table \ref{tab2}. In all cases
but one, the best-fit solution is obtained for normal templates rather
than peculiar templates. The significance of this result is evaluated
by means of the F-test values: the normal solution is strongly favored
(low F-test) for SN~2002lj, SN~2002lp and SN~2002lr, three supernovae
for which features are well fitted both in the red and the blue part
of the spectra. The normal solution is only weakly favored for
SN~2002lq as the relatively featureless red part of the spectrum is
not very discriminant. 

Only in the case of SN~2002li is a solution for an overluminous \Ia
template (SN~1999aa; \citet{Garavini04}) weakly favored over a normal
solution. Given the absence of characteristic features in the red part
of this spectrum and a fair contamination of the spectral signal by
the host light, the F-test does not clearly favor one solution over
the other. At this stage, we consider that SN~2002li is likely a
normal type Ia but this conclusion is not as secure as for the other
\Ia of the sample.

Finally, we note that the best-fit solutions obtained with peculiar
templates are all for overluminous objects, except for SN~2002lp for
which the underluminous SN~1999by is found as the best-fit template.
This might be linked to the strong silicon 5972$ \AA\ $ absorption
observed in this spectrum which is stronger than what is usually
observed in normal \Ia.

\subsubsection{\Ia Phases}

\begin{figure*}
 \resizebox{16cm}{!}{\includegraphics{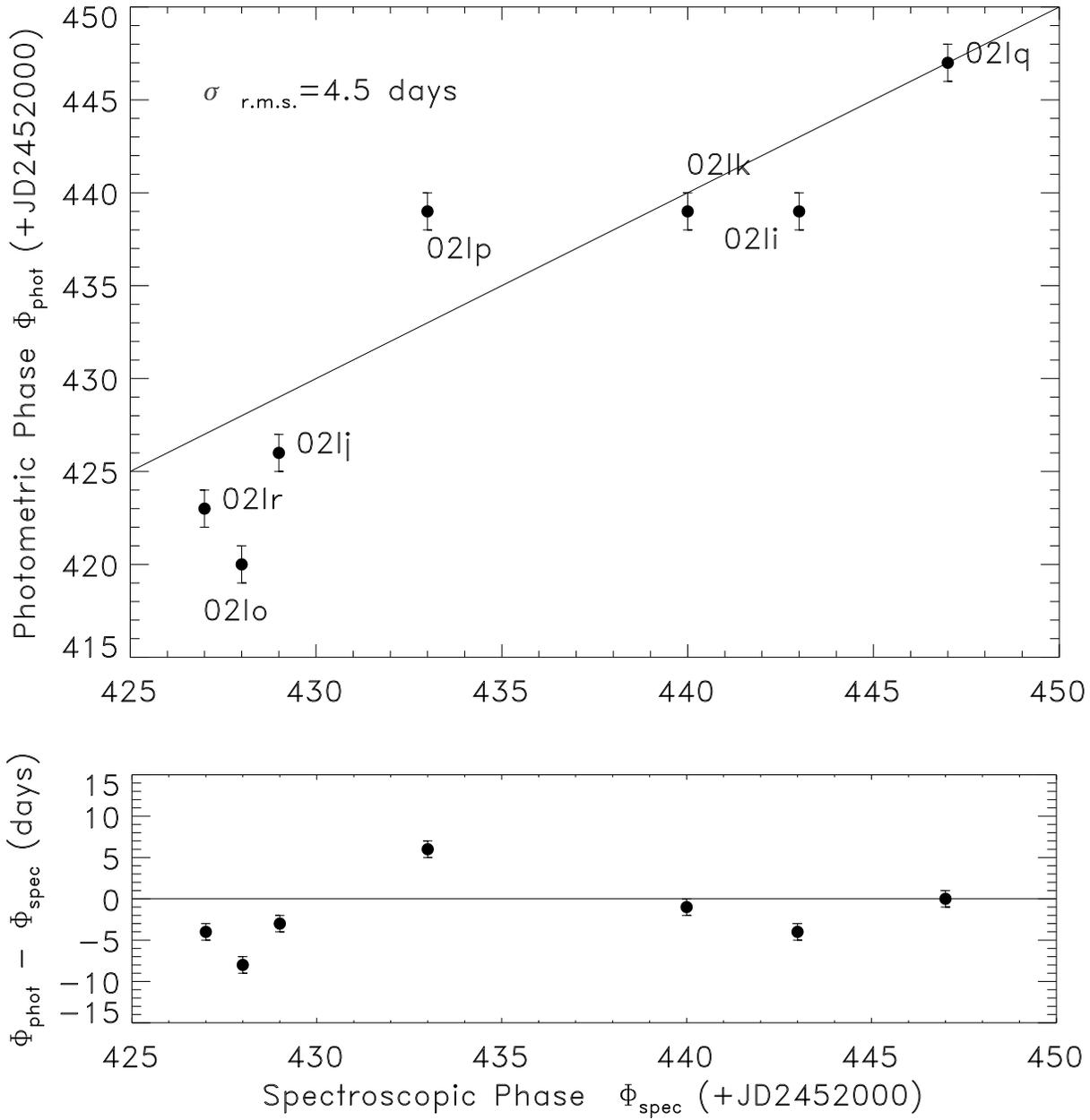}}
\caption{Photometric phase relative to spectroscopic phase for the
  five photometrically followed-up \Ia of our sample (top panel), and
  residuals (bottom panel).  Errors shown are for the photometric
  phase only and assume a conservative $\pm 1$ day error. The
  photometric phase is from fit of the restframe B-band light-curve
  and the spectroscopic phase from $\chi^2$ minimization fitting of
  observed spectra with \Ia spectral templates.}
\label{fig14}
\end{figure*}

An independent way to check the spectroscopic phase is to compare it
with the phase derived from light-curve fitting. We have obtained this
photometric phase as a preliminary step in the analysis of the
photometric follow-up of our supernovae. The photometric analysis,
including photometric calibration and control of systematics will be
published elsewhere. We use here the dates of maximum light as these
are very robust against slight modifications of the photometric
calibration. Table \ref{tab5} gives the photometric date of B-band
maximum (column 2) along with the one inferred from the spectroscopic
analysis (column 3). We conservatively estimate a $\pm 1$ day
uncertainty on the photometric dates.  In Fig. \ref{fig14}, we plot
the photometric Julian date of maximum as a function of the Julian
spectroscopic maximum (top panel). Residuals are shown on the bottom
panel. The dispersion obtained is $\sigma \approx 4.5$ days, slightly
larger than our spectroscopic estimate. It grows up to $\approx 12$
days when considering the phases obtained from the fits with peculiar
 \Ia templates (this difference is mainly due to
SN~2002lr).

\subsection{Host results}

\begin{figure*}
  \resizebox{20cm}{!}{\includegraphics{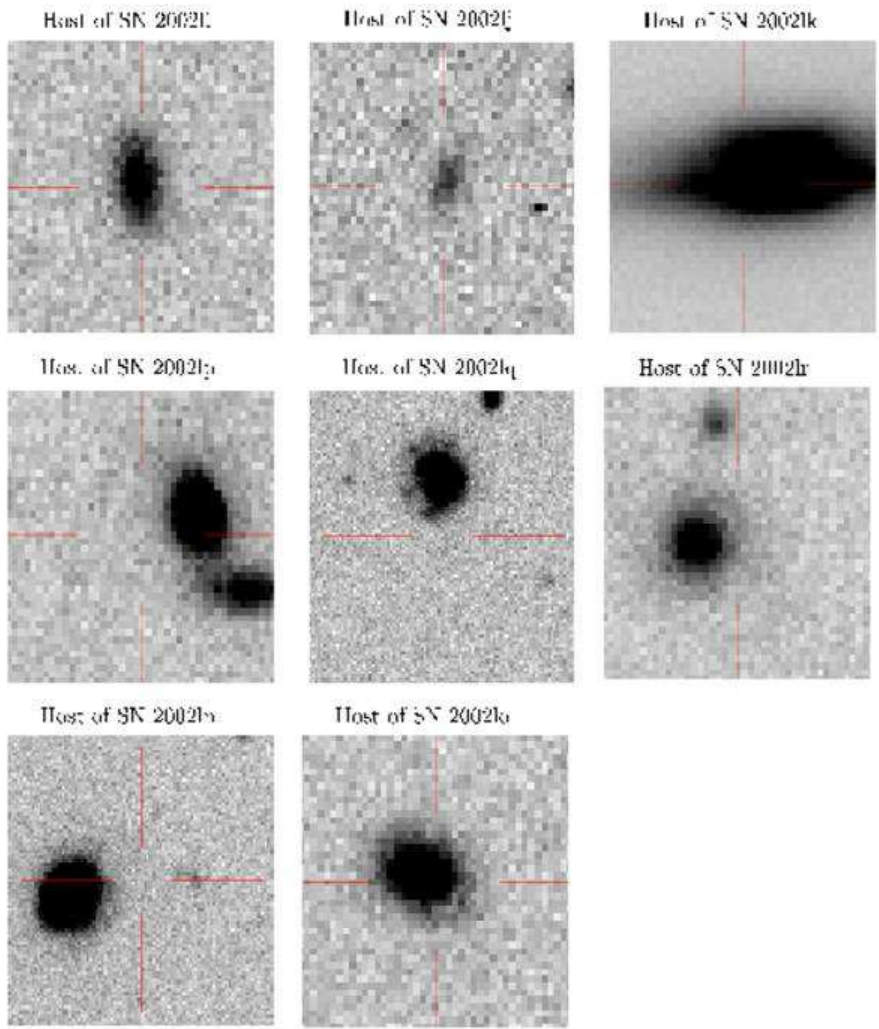} }
\caption{WFS reference images of the host galaxies. Each vignette is a 0.25$\times $0.25 square-arcmin cutout of the full CCD, centered  on the position where the SN exploded (indicated by a cross), except for SN~2002ln and SN~2002lq for which the vignette is 0.5$\times 0.5$ square-arcmin cutout.}
\label{fig15}
\end{figure*}

Figure \ref{fig15} shows WFS reference images of host galaxies of the
eight supernovae. All vignettes are a 0.25$\times $0.25 square-arcmin
g' image centered on the location of the supernova explosion, except
for SN2002ln and SN2002lq for which a 0.5$\times $0.5 square-arcmin
image is shown.

Table \ref{tab6} summarizes the host classification for each
candidate. The lines used for identification are detailed in column 2,
specific comments are given in column 3, and the identification based
on the spectra is presented in column 4. For comparison, the galaxy of
the best-fit model found by ${\cal SN}$-fit is given in column 5. In
general, good agreement is found between the identification drawn from
galaxy features and the galaxy type used to build the best-fit model.

We have also computed host colors from their magnitudes found in the
SDSS online archive.  This was possible for six out of the eight
supernovae of our sample, with the exception of SN~2002lj and
SN~2002lk. Results are given in Table \ref{tab7}. u'-g' and g'-r'
colors have been computed directly from the magnitudes, whereas B-V
colors were calculated using \citet{Fukugita96} color equations. We
compare these colors to the colors of \citet{Frei94} that were
computed using the galaxy energy distributions compiled for each
Hubble type by \citet{Coleman80}. For our purposes, we interpolate the
$z=0, 0.1, 0.4$ and $0.6$ colors of Frei \& Gunn at the redshifts of
our hosts, for each Hubble type. To connect their galaxy types with
ours, we associate their type E to our type 0, Sbc to our type 1 and
Scd and Im to our type 2. We assume that a Sa galaxy (type 1 in our
classification) has colors falling in between their E and Sbc types.
The resulting classification is indicated in column 6. Identification
is possible when at least two out of the three computed colors are
consistent with the given type. We note that in 3 cases, the 3 colors
agree. We compare this color-based type to the spectral type presented
in Table \ref{tab6}. We find general agreement between the various
identifications, except for SN~2002lo for which the color based
identification yields a type 2 whereas the spectroscopic
identification is for a type 1 host.

\subsection{The case of SN~2002lk}

The presence of a strong \ion{Na}{iD} doublet ($\lambda\lambda
5890-5896\AA)$ absorption line in the spectra of SN~2002lk allows for
a determination of the color excess $E(B-V)$ for this object.
Unfortunately, at the redshift of SN~2002lk, the \ion{Na}{iD} line
falls at $\sim 6100\AA$ (in the observer frame; see Fig. \ref{fig7}
and \ref{fig8}), that is at the junction between the blue and the red
part of the spectrum. Hence, the measurement of the equivalent width
of the \ion{Na}{iD} line is quite uncertain ($EW=3\pm 1\AA$), as
this part of the spectrum is most sensitive to possible errors in flux
calibration. Using \citet{Barbon90} and \citet{Turatto03}
empirical correlation between equivalent width and color excess, we
estimate $E(B-V)=0.5 - 1.0$.  We note that these high values, mostly
due to dust absorption in the dense lane visible on the detection
images of SN~2002lk, are consistent with the high value obtained by
\citet{Phillips99} for SN~1986G. The Galaxy contribution to the
reddening is negligeable with $E(B-V) \approx 0.007$ mag.
\citep{Schlegel98}. We have dereddened the June 10th spectrum of
SN~2002lk using various $E(B-V)$ ranging from 0.5 to 1.2 with 0.1
increase step, and for each dereddened spectrum we have searched for
the best-fit template using ${\cal SN}$-fit. The best-fit is obtained
for $E(B-V)=0.9$ (in fair agreement with a preliminary fit to the
  light-curve of SN~2002lk yielding $E(B-V)\sim 0.8$) and has been
presented in Fig. \ref{fig8}. This is well matched by an underluminous
\Ia template a few days before maximum. No satisfactory solution is
obtained when the same dereddened spectrum is fitted with Nobili's
template around maximum. We need to go to higher phase ($\sim 10$
days) to get solution with $\chi^2$ values comparable with the
underluminous solution. However, in this case, the phase is no longer
consistent with the photometric phase of this SN.

\section{Discussion and concluding remarks}
 
With the SDSS-II results to come, the sample of \Ia at intermediate
redshift will increase by an order of magnitude. At present, the \Ia
in this redshift range amount to approximately 35. Our two searches in
1999 and 2002 represent half of them. Indeed, excluding the peculiar
SN~2002lk and the two type-II of the present sample, the set of
spectra presented in this paper, combined with the spectra of
\citet{Balland06}, provides a total of seventeen spectra of normal \Ia
at intermediate redshift. This total sample constitutes a homogeneous
set of spectra in terms of signal-to-noise ratio, wavelength coverage
and resolution. Phases sample the range $-$11 to +9 days\footnote{Except
  for SN~1999dr found at 24 days past maximum.} in a regular way, with
a majority around maximum light, and redshifts range between
$0.15$ and $0.5$. For the 2002 \Ia alone, we classify 20 \% of the
hosts as type 0, 60\% as type 1 and 20\% as type 2. These numbers are
very similar to those obtained for the 1999 \Ia sample (18, 58 and 25\%
respectively). Put altogether, we find 18\% of type 0, 59\% of type 1
and 23\% of type 2. About 82\% of the hosts of our total sample are
spirals, in good agreement with the results of \citet{Sullivan03} for
their low redshift sample: they find 12\%(0), 56\%(1) and 32\%(2),
that is 88\% of spirals for \Ia at $z<0.01$.

We have searched for peculiarities in the spectra of the seventeen \Ia
of the total sample. Although a full computation of the rate of
peculiar objects at intermediate redshift would imply an assessment of
the detection efficiency of such objects in our surveys through
detailed Monte-Carlo simulations, we can draw a first conclusion from
the analyses of this paper and of \citet{Balland06}. We found 0$\pm1$
peculiar objects in the 1999 sample. The same result is obtained in the
2002 sample as the F-test can not discriminate between an SN~1999aa
and a normal best-fit solution for SN~2002li. Thus, for our total
sample, two might be spectrally overluminous, leading to an upper
limit estimate of 12$\%$ of the full sample, perfectly consistent with
the 12\% found by \citet{Li01a} in Monte-Carlo simulations for a
magnitude-limited survey with a 20 day baseline such as ours, with an
extra R-band extinction for overluminous objects of 0.8 mag and an
age-bias cutoff of 7 days.  This however contrasts with the 36\% of
peculiar objects found in local samples \citep{Li01b}. In high
redshift samples, no peculiar objects have been reported
\citep{Perlmutter99,Riess98b}. It is interesting that our estimated
percentage of peculiar objects at intermediate redshift falls in
between the values for low- and high-redshift samples, as it suggests
a continuous sequence in the rate of peculiar \Ia with redshift.

This paper has presented spectra of five Branch-normal \Ia and of two
type II-P supernovae, all at intermediate redshift, observed as part
of an international time program on "{\em $\Omega$ and $\Lambda$ from
  Supernovae, and the Physics of Supernova Explosions}", performed in
May and June 2002. In addition, three spectra of an underluminous
supernova at low redshift have been taken during two consecutive
nights in June 2002. This supernova (SN~2002lk) is interesting, in
addition to being underluminous, it sits in a dust lane of
its host galaxy and it is consequently very reddened (E(B-V)$\approx$
0.9). It closely ressembles SN~1986G, and spectra of this latter
supernova around maximum provide a reasonable fit to SN~2002lk. Once
dereddened, the spectra of SN~2002lk are well fitted by SN~1991bg-like
supernovae. We conclude that, from a spectroscopic point of view,
SN~2002lk belongs to this class of underluminous supernovae.

Regarding \Ia alone, the new set presented in this paper, added to the
spectra of \citet{Balland06}, provides more than fifteen spectra of
supernovae in the still largely unexplored intermediate redshift range
$0.15 \lesim z \lesim 0.35$. In this paper, we have concentrated on
some physical properties of the \Ia in this redshift range.

Regarding the phase and redshift distribution, our global sample is
representative of a population of type Ia supernovae around
maximum light. The use of both the blue and red arms of the ISIS
spectrometer for the present sample gives access to the UV part of
spectra down to $3000\AA\ $. This was also the case for the two April
spectra in 1999 \citep{Balland06}. This part of the spectrum is
expected to be sensitive to metallicity and will be interesting to
study in larger scale samples.

As far as host types are concerned, the analysis of this new set of
\Ia confirms the fractions of Hubble types of host populations found
in other samples at lower or higher redshift. This conclusion, that
reinforces the idea that no systematic difference exists in supernova
populations at different redshifts was not so clear in our 1999
sample. The inclusion of this new set in our analysis allows us to
conclude more firmly on this matter.

Finally, our analysis reveals that very few if any of these supernovae
at intermediate redshift show signatures of peculiarities, contrary to
what is observed in low-redshift samples. In particular, no
overluminous supernova is found, although we have undertaken a
specific search of these supernovae, both by fitting our data with
peculiar spectral templates and by a close examination of the
ejection velocities of some spectral features that are a potential
signature of peculiarity. This is somewhat contradictory with what one
could have expected, as overluminous supernovae are often found in
regions of active star formation \citep{Li01a}, more numerous at high
redshift.  Although based on a still low number sample, this fact is
an important conclusion regarding the potential contamination of the
Hubble diagram by such objects.

\begin{acknowledgements}
  The observations described in this paper were obtained, as part of
  the International Time Programme on {\em $\Omega$ and $\Lambda$ from
    Supernovae, and the Physics of Supernova Explosions}, as
  visiting/guest astronomers at the INT and WHT, operated by the Royal
  Greenwich Observatory at the Spanish Observatorio del Roque de los
  Muchachos of the Instituto de Astrofisica de Canarias. We thank the
  dedicated staffs of these observatories for their assistance. We
  also thank A.-L. Huat for her participation in reducing the data
  during her master training period. S. Fabbro, A.-M. Mour\~ao and R.
  Pascoal thank the support by Funda\c{c}\~ao para a Ci\^{e}ncia e
  Tecnologia under project POCTI/FNU/43749/2001.

\end{acknowledgements}

\bibliographystyle{aa}
\bibliography{bibi}

\newpage

\vfill
\eject

\clearpage
\newpage

\newpage

\newpage

\newpage

\newpage

\newpage

\clearpage

\clearpage

\clearpage

\clearpage

\newpage

\small
\begin{longtable}[l]{llccccccc}
\caption[c]{\label{tab1} Log of Spectroscopic Observations (WHT+ISIS)}\\
\hline \hline
SN  & R.A. (2000.0) & Dec. (2000.0) & Date & g' & Date of & JD & Exposure time (s) & Resolution$\ddagger$ \\
& hh mm ss & $^o$ ' " & of Discovery  (2002)& mag$\dagger$ & Spectroscopy & +2400000 & Blue/Red & {\bf ($\AA$)}\\
\hline \endfirsthead
2002li  &       15:59:03.08 & +54:18:16.0 & Jun. 4 & 20.6& Jun. 10      & 52436.6  & 3$\times$900 & 8.3\\
2002lj& 16:19:19.65 & +53:09:54.2 & Jun. 5& 19.7&        Jun. 11        & 52437.4 & 2$\times$900 & 7.2\\
2002lk (a) & 16:06:55.92 & +55:28:18.2 & Jun. 6 & 18.3 & Jun. 10        & 52436.4  & 300+600 & 8.4\\
2002lk (b)& & & & & Jun. 11                                                     & 52437.4  & 600 & 8.2\\
2002ln& 16:39:24.93 & 41:47:29.0 & Jun. 6 & 22.5 &      Jun. 10 & 52436.6 & 2$\times$900 & 8.4 \\
2002lo& 16:39:56.42 & +42:19:20.5 & Jun. 6 & 21.9 & Jun. 11     & 52437.6  & 2$\times$900 & 7.1\\
2002lp& 16:40:11.45 & +42:28:30.2 & Jun. 6 & 20.4 & Jun. 10     & 52436.5  & 2$\times$900 & 8.3\\
2002lq& 16:40:28.83 & +41:14:09.1 & Jun. 6 & 21.9 & Jun. 10     & 52436.4  & 2$\times$900 & 8.4\\
2002lr& 22:33:12.59 & +01:05:56.7 & Jun. 6 & 20.8 & Jun. 10     & 52436.7  &  2$\times$900 & 8.3\\
\hline
\end{longtable}

\footnotesize{Notes:}\\
\hspace*{5mm}\footnotesize{$^\dagger$ Approximate magnitude at discovery}\\
\hspace*{5mm}\footnotesize{$^\ddagger$ Full width at half maximum of [OI] night-sky line at 5577$\AA$}\\

\small
\begin{longtable}[c]{lccccccc}
\caption[c]{\label{tab2} Results from ${\cal SN}$-fit}\\
\hline \hline
SN     & Best fit SN& \% (host)$^\dagger$ & $z_f$  & $z_h$  & Spectroscopic & Reduced & d.o.f.$^\ast$ \\
name &  template (reference) & & & & phase$^\ddagger$ & $\chi^2$ &\\ 
\hline
2002li & 03du (Anupama \etal 2005) & 50(Sa) & 0.327 & 0.328 & -7 & 1.21 & 827\\
2002lj & 92A (Kirshner \etal 1993) & 13(Sa) & 0.181& 0.183& +7& 2.57& 1217\\
2002lk (Jun. 10)& 86G (Phillips \etal 1987) & 35(S0) & 0.023& 0.033 & +2 & 6.83& 999\\
2002lk (Jun. 11)& 86G (Phillips \etal 1987) & 26(S0) & 0.024& 0.033 & +2 & 3.26& 954\\
2002lk (dereddened)$^{\ast\ast}$& 99by (Garnavich \etal 2004) & 19(S0) & 0.022 & 0.033 & -4 & 8.16 & 1205\\ 
2002ln & 99em (Hamuy \etal 2001) & 38(Sb) & 0.144 & 0.143 & +23$^{\ast\ast\ast}$ & 0.954 & 1486\\
2002lo &99em (Hamuy \etal 2001)&55(Sa) &  0.144 & 0.136 & +13$^{\ast\ast\ast}$ & 1.29 & 1030\\
2002lp & 92A (Kirshner \etal 1993) & 46(S0) & 0.146 & 0.137 & +3 & 2.59 & 1135\\
2002lq & 94D (Patat \etal 1996) & 37(Sa) & 0.275 & 0.274 & -11 & 0.81 & 859\\
2002lr & 92A (Kirshner \etal 1993) & 42(Sa) & 0.256 & 0.258 & +9 & 0.83& 908\\
\hline
\end{longtable}

\footnotesize{Notes:}\\
\hspace*{5mm}\footnotesize{$^\dagger$ \%(host) = $\beta/(\alpha + \beta)\times 100$ (see text)}\\
\hspace*{5mm}\footnotesize{$^\ddagger$ Typical error is $\pm 3$ days}\\
\hspace*{5mm}\footnotesize{$^\ast$ Number of degrees of freedom}\\
\hspace*{5mm}\footnotesize{$^{\ast\ast}$ Spectrum dereddened using Howarth (1983) law and E(B-V)=0.9; see text}\\
\hspace*{5mm}\footnotesize{$^{\ast\ast\ast}$ With respect to explosion date}\\

\small

\begin{longtable}[c]{lccc}
\caption{\label{tab3} SN expansion velocities}\\

\hline
\hline
SN Name & Phase &$v_{\mathrm{Ca\,{\sc II}}}^a$ & $v_{\mathrm{Si\,{\sc II}}}^b$ \\ \hline
2002li & -7 &-16900& $\sim$ -13000\\
2002lj & +7 & -13300 & -10000 \\
2002lk (Jun. 10) & +2 & & -13550 \\
2002lp & +3 & -13800  &  -10400 \\
2002lq & -11 & -22950 &\\
2002lr & +9 & -12200  & -8600 \\
\hline
\end{longtable}

\footnotesize{Notes:}\\
\hspace*{5mm}\footnotesize{$^a$ Ca H\& K $\lambda 3945$ in km/s; typical error is 500 km/s}\\
\hspace*{5mm}\footnotesize{$^b$ Si\,{\sc ii} $\lambda 6355$ in km/s; typical error is 200 km/s}\\

\small
\begin{longtable}[c]{ccccccccc}
\caption[]{\label{tab4} Results from${\cal SN}$-fit using peculiar \Ia templates}\\
\hline\hline SN  & \multicolumn{4}{c}{Peculiar} &
\multicolumn{3}{c}{Normal} &
F-test\\
&  \multicolumn{4}{c}{------------------------------------------} & \multicolumn{3}{c}{-----------------------------} &probability (\%) \\
&  Best-fit & Phase & $\chi^2$ & d.o.f. & Phase & $\chi^2$ & d.o.f. & \\
\hline
2002li & 99aa$^1$ & -7 & 1.16 & 811 & -7  & 1.21 & 827 & 0.34\\
2002lj & 99aa & +14 & 2.99 & 1276 & +7 & 2.57 & 1217 & 1.1$\times 10^{-6}$\\
2002lp & 99by$^2$ & -5& 3.38 & 1251 & +3 & 2.59 & 1135 & 4.4$\times 10^{-15}$\\
2002lq & 99aa & -7 & 0.84 & 896 & -11 & 0.81& 859 & 0.30\\
2002lr & 91T$^3$ & -12 & 1.03 & 988 & +9 & 0.83 & 908  & 6.2$\times 10^{-9}$\\
\hline
\end{longtable}

\footnotesize{Notes:}\\
\hspace*{5mm}\footnotesize{$^1$ Spectral template from Garavini et al. (2004)}\\
\hspace*{5mm}\footnotesize{$^2$ Spectral template from Garnavich et al.(2004) }\\
\hspace*{5mm}\footnotesize{$^3$ Spectral template from Mazzali et al. (1995)}\\

\small
\begin{longtable}[c]{lcc}
\caption{\label{tab5} Spectroscopic and light-curve dates of maximum for the five photometrically followed-up \Ia}\\
\hline\hline
SN & Date of B max. (restframe)& Maximum date$\dagger$\\ 
\hline
2002li & 13/06 & 17/06\\ 
2002lj & 31/05 & 03/06\\ 
2002lk (Jun. 10) & 13/06 & 14/06 \\
2002lo & 25/05 & 02/06 \\
2002lp & 13/06 & 07/06 \\
2002lq &  21/06 & 21/06 \\
2002lr & 28/05 & 01/06  \\ 
\hline
\end{longtable}

\footnotesize{Notes:}\\
\hspace*{5mm}{\footnotesize $\dagger$ Date estimated from the date of
  spectroscopic observation corrected for the best-fit spectroscopic phase}\\

\small

\begin{longtable}[c]{ccccc}
\caption{\label{tab6}SN host spectral classification.}\\
\hline
\hline
SN & Indicators$^\ast$ &  Comments & Host Spectral 
 & ${\cal SN}$-fit  \\ 
 &  &  & type$^\dagger$ &  \\
\hline
2002li & s. [OII], s. [OIII], s. B4000, Ca H\&K, s. H$\beta$, NaD  & early spiral & 1 & Sa\\
2002lj & w. [OII], w. H$\beta$, $H\alpha$ ? & very weak gal. features & 1 ? & Sa\\
2002lk & s. NaD, H$\alpha$, [NII] & early spiral on detection images  & 1 & S0\\
2002ln & H$\alpha$& very few features & 1 & Sb \\
2002lo & H$\alpha$, H$\beta$, H$\gamma$, H$\delta$, [OII], [OIII], NaD & spiral & 1& Sa\\
2002lp & H$\alpha$?, w. [OII], w. [OIII] & weak features & 0 & S0\\
2002lq & H$\alpha$, H$\beta$, s. [OII], s. [OIII] & early spiral & 2 & Sa\\
2002lr & [OII], [OIII], H$\alpha$ ? & spiral & 1 & Sa\\
\hline
\end{longtable}

\footnotesize{Notes:}\\
\hspace*{5mm}\footnotesize{$^\dagger$ Type 0: spheroidal (E/S0); Type 1: early-type spiral (Sa/Sb); Type 2: late-type spiral (Sc/Starburst)}\\
\hspace*{5mm}\footnotesize{$^\ast$ B4000=4000\AA $\ $ break, s.=strong, w.=weak}\\

\small
\begin{longtable}[c]{cccccc}
\caption{\label{tab7} Host galaxy SDSS colors corrected for Milky-Way extinction}$^{\dagger}$\\
\hline\hline
SN & E(B-V) & B-V & g'-r' & u'-g' & Color-based host type\\
\hline

2002li &0.011& 1.09& 0.91 & 0.69 & 1\\
2002ln & 0.008 & 1.55 & 1.39 & -0.55 & 1\\
2002lo & 0.011& 0.70 & 0.50 & 0.98 & 2\\
2002lp & 0.011 & 1.08 & 0.90 & 1.75 & 0\\
2002lq & 0.008& 0.93 & 0.75 & 0.92 & 2\\
2002lr & 0.09 & 1.07 & 0.89 & 0.72 & 1\\
\hline
\end{longtable}

\footnotesize{Note:}\\
\hspace*{5mm}\footnotesize{Computations based on magnitudes from SDSS query page. No information on SN~2002lj or SN~2002lk has been found.}

\end{document}